\documentclass[pre,tighten,showpacs]{revtex4-1}

\usepackage{amssymb,amsmath}

\usepackage{graphicx}

\usepackage{color}

\newcommand\beq{\begin{equation}}
\newcommand\eeq{\end{equation}}
\newcommand\beqa{\begin{eqnarray}}
\newcommand\eeqa{\end{eqnarray}}
\newcommand{\dd}{\text{d}}

\newcommand{\al}{\alpha}

\newcommand{\nuzt}{\nu_{0|2}}
\newcommand{\nuto}{\nu_{2|1}}

\newcommand{\nufz}{\nu_{4|0}}

\begin{document}

\title{Inelastic Maxwell models for monodisperse gas-solid flows}
\author{Aleksander Kubicki}
\affiliation{Departamento de F\'{\i}sica, Universidad de Extremadura, Universidad de Extremadura, E-06071 Badajoz, Spain}
\author{Vicente Garz\'o}
\email{vicenteg@unex.es} \homepage{http://www.eweb.unex.es/eweb/fisteor/vicente/}
\affiliation{Departamento de F\'{\i}sica and Instituto de Computaci\'on Cient\'{\i}fica Avanzada (ICCAEx), Universidad de Extremadura, E-06071 Badajoz, Spain}

\begin{abstract}

The Boltzmann equation for $d$-dimensional inelastic Maxwell models is considered to analyze transport properties for monodisperse gas-solid suspensions. The influence of the interstitial gas phase on the dynamics of solid particles is modeled via a viscous drag force. The Chapman-Enskog method is applied to solve the inelastic Boltzmann equation to first order in the deviations of the hydrodynamic fields from their values in the homogeneous cooling state. Explicit expressions for the Navier-Stokes  transport coefficients are \emph{exactly} obtained in terms of both the coefficient of restitution and the friction coefficient characterizing the amplitude of the external force.  The conditions under which a hydrodynamic regime independent of the initial conditions is reached are widely discussed. Finally, the results derived here are compared with those previously obtained for inelastic hard spheres in steady state conditions by using the so-called first Sonine approximation.
\end{abstract}

\draft
\date{\today}
\maketitle

\section{Introduction}
\label{sec1}

Granular gases are usually modeled as a gas of hard spheres whose collisions are inelastic (inelastic hard spheres, IHS). In the simplest model, the spheres are completely smooth and the inelasticity of collisions is characterized by a (positive) constant coefficient of normal restitution $\al\leq 1$. The case $\al=1$ corresponds to elastic collisions (ordinary gases). In addition, although in nature granular matter is usually surrounded by an interstitial fluid (like air), the effect of the latter on the dynamic properties of solid particles is usually neglected in most of the theoretical works.
However, it is known that the influence of the interstitial gas phase on solid particles can be important in a wide range of practical applications and physical phenomena, like for instance species segregation \cite{segregation}.  Needless to say, at a kinetic theory level, the description of rapid gas-solid flows is an intricate problem since it involves two phases and hence, one would need to solve a set of two coupled Boltzmann kinetic equations for each one of the different phases. In order to gain some insight into this problem, an usual approach \cite{K90,KH01} is to consider a single Boltzmann equation for the solid particles where the effect of the gas phase on the latter ones is incorporated by means of an effective external force.

Recently, a model for a monodisperse gas-solid suspension described by the Enskog kinetic theory (and hence, it applies to moderate densities) has been proposed \cite{GTSH12}. Unlike previous efforts for similar suspensions, the gas phase contribution to the instantaneous acceleration appearing in the Enskog equation is modeled through a Langevin like term. Although the model can be in principle applied to a wide parameter space (e.g., high Reynolds numbers), the theory \cite{GTSH12} was limited to low Reynolds number flow. The model proposed in Ref.\ \cite{GTSH12} presents some similarities with a model widely used by Puglisi and co-workers \cite{andrea,GMV13} in computer simulations to homogeneously fluidize a granular gas by an external driving force. The use of this sort of ``thermostats'' is very common in simulations as a way to inject energy into the system and reach stationary states. More specifically, the external force employed in Refs.\ \cite{GTSH12,GMV13,andrea} is composed by two terms: (i) a drag force proportional to the velocity of the particle and (ii) a stochastic force (Langevin model) with the form of a Gaussian white noise where the particles are randomly kicked between collisions \cite{WM96}. While the first term tries to mimic the friction of grains with a viscous interstitial fluid, the second term attempts to model the energy transfer from the surrounding fluid to granular particles. It must be noted that while the friction coefficient associated with the drag force and the amplitude of the stochastic force of the model proposed in Ref.\ \cite{andrea} are related in the same way as in the well-known fluctuation-dissipation theorem \cite{K92} of molecular gases, those coefficients are independent in the model of Ref.\ \cite{GTSH12} since they are defined in terms of parameters such as the Reynolds number, the volume fraction and the ratio of the densities of the solid and gas phases. This is the main difference between the models introduced in Refs.\ \cite{GTSH12} and \cite{andrea}. In particular, when the mean flow velocities of solid and gas phases coincide, then the coefficient associated with the Langevin-like term vanishes (see Eq.\ (8.2) of Ref.\ \cite{GTSH12}) and the presence of the interstitial fluid is only accounted for by the external drag force.

On the other hand, even for the dry granular case (i.e., when the gas phase effects over grains are neglected) \cite{BDKS98,GD99}, the forms of the Navier-Stokes transport coefficients of IHS  cannot be \emph{exactly} obtained \cite{GTSH12,GMV13} and hence, one has to consider additional approximations such as the truncation of a Sonine polynomial expansion. A possible way of circumventing the technical difficulties associated with the complex mathematical structure of the (linearized) Enskog-Boltzmann collision operator for IHS is to consider the so-called inelastic Maxwell models (IMM), namely, models where the collision rate is independent of the relative velocity of the two colliding spheres. The use of IMM allows one to get in a clean way and without any uncontrolled approximation the dependence of the transport coefficients on the coefficient of restitution \cite{S03}. Very recently \cite{MGV14}, the Boltzmann kinetic equation for a driven granular gas of IMM has been solved by means of the Chapman-Enskog method \cite{CC70}. As in previous works \cite{GMV13,andrea}, the gas was fluidized by a thermostat composed by both the drag and stochastic terms. In addition, for the sake of simplicity, the coefficients associated with both forces were not considered as independent parameters. However, in spite of the above simplification, the evaluation of the transport coefficients in the driven case for general \emph{unsteady} states requires to numerically solve a set of differential equations and hence, only exact expressions were derived under steady state conditions \cite{MGV14}.

In this paper, we consider a simplified version of the model of suspensions used in Refs.\ \cite{GTSH12,GMV13,MGV14} where only the drag force term is accounted for. As mentioned before, this situation could correspond to a gas-solid flow where the mean velocity of the particles follows the velocity of the fluid (such as in the case of the simple shear flow \cite{TK95}). It must be remarked that the above drag force model has been recently considered in different papers \cite{H13,SMMD13,WGZS14} to study the shear rheology of frictional hard-sphere suspensions. The use of this drag model allows one to get exact results for the transport coefficients for general unsteady conditions.

The main advantage of using IMM instead of IHS is that a collision moment of order $k$ of the Boltzmann collision operator can be exactly expressed in terms of moments of order less than or equal to $k$ \cite{TM80,GS03}. These collisional moments are proportional to an effective collision frequency $\nu_0$, which in principle can be freely chosen. As in previous works \cite{SG07}, we will consider here two classes of IMM: (a) a collision frequency $\nu_0$ independent of temperature (Model A) and (b) a collision frequency $\nu_0(T)$ monotonically increasing with temperature, namely, $\nu_0\propto n T^q$ (Model B). While Model A is closer to the original model of Maxwell molecules for elastic gases \cite{TM80,GS03}, Model B (with $q=\frac{1}{2}$) is closer to IHS. The possibility of considering a general function $\nu_0(T)$ is akin to the class of inelastic repulsive models introduced by Ernst and co-workers \cite{Ernst}.

The plan of the paper is as follows. In section \ref{sec2}, the Boltzmann equation for IMM of granular gases driven by an external drag force is introduced and the explicit expressions of the collisional moments needed to get the transport coefficients are given. Section \ref{sec3} addresses the study of the so-called homogeneous cooling state (HCS) where a scaling solution is proposed that depends on granular temperature $T$ only through the dimensionless velocity $\mathbf{c}=\mathbf{v}/v_0(T)$ ($v_0(T)=\sqrt{2T/m}$ being the thermal velocity). This solution is similar to the one obtained in previous works on dry granular gases \cite{NE98}. The Chapman-Enskog expansion around the \emph{local} version of the HCS is described in section \ref{sec4} while the expressions of the Navier-Stokes transport coefficients $\eta$ (shear viscosity), $\kappa$ (thermal conductivity) and $\mu$ (not present for elastic collisions) are determined in section \ref{sec5}. The dependence of the above transport coefficients on the parameter space of the problem is analyzed and compared with results of IHS in the case of low Reynolds numbers and for steady states in section \ref{sec6}. The paper is closed in section \ref{sec8} with some conclusions.

\section{Boltzmann kinetic theory for inelastic Maxwell models of driven granular gases}
\label{sec2}

Let us consider a granular fluid modeled as an inelastic Maxwell gas of hard disks ($d=2$) or spheres ($d=3$). The inelasticity of collisions among all pairs is accounted for by a {\em constant} (positive) coefficient of restitution $\alpha \leq 1$ that only affects the translational degrees of freedom of grains. As said in the Introduction, in order to assess the effects of the interstitial fluid on particles, an external nonconservative force is incorporated into the corresponding kinetic equation of the solid particles. Under these conditions, in the low-density regime, the one-particle velocity distribution function $f({\bf r}, {\bf v}, t)$ of grains obeys the \emph{inelastic} Boltzmann equation
\begin{equation}
\label{2}
\frac{\partial f}{\partial t}+{\bf v}\cdot \nabla f+{\cal F} f=J[\mathbf{v}|f,f],
\end{equation}
where
\beq
\label{3}
J\left[{\bf v}_{1}|f,f\right] =\frac{\nu}{n\Omega_d} \int \; \dd{\bf v}_{2}\int
\dd\widehat{\boldsymbol{\sigma}} \left[ \alpha^{-1}f({\bf v}_{1}')f({\bf v}_{2}')-
f({\bf v}_{1})f({\bf v}_{2})\right] \;
\end{equation}
is the Boltzmann collision operator for IMM. Here,
\begin{equation}
\label{4}
n=\int \; \dd{\bf v}f({\bf v})
\end{equation}
is the number density, $\nu$ is an effective collision frequency assumed to be independent of the coefficient of restitution $\al$, $\Omega_d=2\pi^{d/2}/\Gamma(d/2)$ is the total solid angle in $d$ dimensions and
$\widehat{\boldsymbol{\sigma}}$ is a unit vector along the line of
the two colliding spheres. In addition, the primes on the velocities denote the initial values $\{{\bf
v}_{1}^{\prime}, {\bf v}_{2}^{\prime}\}$ that lead to $\{{\bf v}_{1},{\bf v}_{2}\}$ following a binary collision:
\begin{equation}
\label{5}
{\bf v}_{1}^{\prime}={\bf v}_{1}-\frac{1}{2}\left(
1+\alpha ^{-1}\right)(\widehat{\boldsymbol{\sigma}}\cdot {\bf
g}_{12})\widehat{\boldsymbol {\sigma}}, \quad {\bf v}_{2}^{\prime}={\bf
v}_{2}+\frac{1}{2}\left( 1+\alpha^{-1}\right)
(\widehat{\boldsymbol{\sigma}}\cdot {\bf
g}_{12})\widehat{\boldsymbol{\sigma}}\;,
\end{equation}
where ${\bf g}_{12}={\bf v}_1-{\bf v}_2$ is the relative velocity of the colliding pair. Moreover, in Eq.\ \eqref{2} ${\cal F}$ is an operator representing the effect of an external force.

A very usual form of the fluid-solid interaction force in high-velocity gas-solid flows is a viscous drag force given by
\beq
\label{5.0}
\mathbf{F}^{\text{drag}}=-m\gamma (\mathbf{v}-\mathbf{U}_g)
\eeq
where $m$ is the mass of a particle, $\mathbf{v}$ is the particle velocity and $\mathbf{U}_g$ is the (known) mean velocity of the interstitial fluid \cite{H13,SMMD13,WGZS14}. The friction coefficient $\gamma$ is proportional to the viscosity $\mu_g$ of the surrounding fluid and will be assumed to be a constant. Thus, according to Eq.\ \eqref{5.0}, the drag force contributes to the Boltzmann equation with a term of the form
\beq
\label{5.1}
{\cal F}f=-\gamma \Delta \mathbf{U}\cdot \frac{\partial f}{\partial \mathbf{V}}-\gamma\frac{\partial}{\partial \mathbf{V}}
\cdot \mathbf{V} f,
\eeq
where $\Delta \mathbf{U}=\mathbf{U}-\mathbf{U}_g$, ${\bf V}\equiv {\bf v}-{\bf U}$ is the peculiar velocity and
\begin{equation}
\label{5.2}
{\bf U}=\frac{1}{n}\int \;
\dd{\bf v} \; {\bf v}\;  f({\bf v})
\end{equation}
is the mean flow velocity of solid particles. The Boltzmann equation \eqref{2} can be more explicitly written when one takes into account the form \eqref{5.1} of the forcing term ${\cal F}f$:
\begin{equation}
\label{5.3}
\frac{\partial f}{\partial t}+{\bf v}\cdot \nabla f-\gamma \Delta \mathbf{U}\cdot \frac{\partial f}{\partial \mathbf{V}}-\gamma\frac{\partial}{\partial \mathbf{V}}
\cdot \mathbf{V} f=J[\mathbf{v}|f,f].
\end{equation}
It is important to remark that when $\Delta \mathbf{U}=\mathbf{0}$, the model proposed in Ref.\ \cite{GTSH12} for monodisperse suspensions reduces in the dilute limit to the Boltzmann equation \eqref{5.3} since the Langevin-like term due to fluid-particle interactions (which is proportional to $\Delta \mathbf{U}$) is zero in this situation \cite{YZHH13}. In this context, the results derived in this paper can be considered of practical interest to analyze linear transport in dilute gas-solid flows when the mean flow velocity of solid and gas phases are the same \cite{TK95}.
Moreover, it has been also shown \cite{L01} that in the case of hard spheres the drag force term $\partial_\mathbf{v}\cdot \mathbf{v} f$ arises from a (logarithmic) change in the time scale of the hard sphere system without external force.

The other relevant hydrodynamic velocity moment of the distribution $f$ is the so-called \emph{granular} temperature. It is defined as
\begin{equation}
\label{6}
T=\frac{m}{d n}\int \; \dd{\bf v}\; V^2\; f({\bf v}).
\end{equation}
The corresponding macroscopic balance equations for density, momentum, and energy follow directly from Eq.\ ({\ref{2}) by multiplying with $1$, $m{\bf v}$, and $\frac{1}{2}mv^2$ and integrating over ${\bf v}$. The result is
\begin{equation}
\label{2.7} D_{t}n+n\nabla \cdot {\bf U}=0\;,
\end{equation}
\begin{equation}
\label{2.8} D_{t}U_i+\rho^{-1}\nabla_j P_{ij}=-\gamma \Delta U_i\;,
\end{equation}
\begin{equation}
\label{2.9} D_{t}T+\frac{2}{dn}\left(\nabla \cdot {\bf
q}+P_{ij}\nabla_j U_i\right) =-(2 \gamma+\zeta) T\;.
\end{equation}
Here, $\rho=m n$ is the mass density, $D_{t}\equiv \partial _{t}+{\bf U}\cdot \nabla$ and the microscopic
expressions for the pressure tensor ${\sf P}$, the heat flux ${\bf
q}$, and the cooling rate $\zeta$ are given, respectively, by
\begin{equation}
{\sf P}=\int \dd{\bf v}\;m{\bf V}{\bf V}\,f({\bf v}),
\label{2.10}
\end{equation}
\begin{equation}
{\bf q}=\int \dd{\bf v}\,\frac{1}{2}m V^{2}{\bf V}\,
f({\bf v}), \label{2.11}
\end{equation}
\begin{equation}
\label{2.12} \zeta=-\frac{1}{d n T}\int\, \dd{\bf v} \; m\; V^2\; J[{\bf v}|f,f].
\end{equation}
The balance equations (\ref{2.7})--(\ref{2.9}) apply regardless of the details of the interaction model considered. The influence of the collision model appears through the $\alpha$-dependence of the cooling rate and of the momentum and heat fluxes.

One of the advantages of the Boltzmann equation for Maxwell models (both elastic and inelastic) is that the collisional moments of the operator $J[f,f]$ can be \emph{exactly} evaluated in terms of the moments of the distribution $f$, without the explicit knowledge of the latter \cite{TM80,GS03}. More explicitly, the collisional moments of order $k$ are given as a bilinear combination of moments of order $k'$ and $k''$ with $0\leq k'+k''\leq k$. In the case of IMM, the collisional moments involved in the calculation of the momentum and heat fluxes as well as in the fourth cumulant are given by \cite{S03,SG07}
\begin{equation}
\label{2.13}
\int\; \dd\mathbf{v}\; m\; V_iV_j\; J[f,f]=-\nu_{0|2}\left(P_{ij}-p\delta_{ij}\right)-\nu_{2|0} p \delta_{ij},
\end{equation}
\begin{equation}
\label{2.14}
\int\; \dd\mathbf{v}\; \frac{m}{2}\;V^2\;\mathbf{V}\, J[f,f]=-\nu_{2|1}\mathbf{q},
\end{equation}
\begin{equation}
\label{2.15}
\int\; \dd\mathbf{v}\; \;V^4\; J[f,f]=-\nu_{4|0}\langle V^4 \rangle+\lambda_1 d^2\frac{pT}{m^2}-
\frac{\lambda_2}{nm^{2}}\left(P_{ij}-p\delta_{ij}\right)\left(P_{ji}-p\delta_{ij}\right).
\end{equation}
Here, $p=nT$ is the hydrostatic pressure,
\beq
\nuzt=\frac{(1+\al)(d+1-\al)}{2d}\nu_0, \quad \nu_{2|0}=\frac{d+2}{4d}(1-\alpha^2)\nu_0,
\label{2.16}
\eeq
\beq
\nuto=\frac{(1+\al)\left[5d+4-\al(d+8)\right]}{8d}\nu_0,
\label{2.17}
\eeq
\beq
\nufz=\frac{(1+\al)\left[12d+9-\alpha(4d+17)+3\alpha^2-3\alpha^3\right]}{16d}\nu_0,
\label{2.18}
\eeq
\beq
\lambda_1=\frac{(d+2)(1+\al)^2\left(4d-1-6\al+3\al^2\right)}{16d^2}\nu_0,
\label{2.19}
\eeq
\beq
\lambda_2=\frac{(1+\al)^2\left(1+6\al-3\al^2\right)}{8d}\nu_0,
\label{2.20}
\eeq
and we have introduced the fourth-degree isotropic velocity moment
\beq
\label{2.21}
\langle V^4 \rangle=\int\; \dd \mathbf{v}\; V^4\; f(\mathbf{v}).
\eeq
In Eqs.\ \eqref{2.16}--\eqref{2.20}, we have called $\nu_0\equiv 2\nu/(d+2)$. According to Eqs.\ \eqref{2.13} and \eqref{2.16}, $\nu_0$ represents the effective collision frequency associated with the shear viscosity of a dilute elastic gas in the absence of the drag force. Moreover, the expression of the cooling rate $\zeta$ of IMM can be exactly obtained from Eq.\ \eqref{2.13}:
\beq
\label{2.22}
\zeta=\frac{d+2}{4d}(1-\al^2)\nu_0.
\eeq

The results \eqref{2.13}--\eqref{2.20} apply regardless of the specific form of the collision frequency $\nu_0$. On physical grounds $\nu_0\propto n$. In the case of \emph{elastic} Maxwell molecules, $\nu_0$ is independent of temperature. However, in order to correctly describe the velocity dependence of the original IHS collision rate, one usually assumes that $\nu_0$ is proportional to $T^q$ with $q=\frac{1}{2}$. Here, as in previous works on IMM \cite{SG07,MGV14}, we take $\nu_0\propto n T^{q}$, with $0\leq q\leq \frac{1}{2}$. The case $q=0$ is closer to the original Maxwell model of elastic particles while the case $q=\frac{1}{2}$ is closer to hard spheres. We will refer here to Model A when $q=0$ while the case $q\neq 0$ will be referred to as Model B.

\section{Homogeneous cooling state}
\label{sec3}

Before analyzing inhomogeneous states, it is quite convenient first to study the HCS problem. In this case, the density $n$ is constant and the time-dependent temperature $T(t)$ is spatially uniform. Moreover, for the sake of simplicity, we also assume that $\mathbf{U}=\mathbf{U}_g=\mathbf{0}$. Consequently, the Boltzmann equation \eqref{2} for the homogeneous distribution $f_h$ becomes
\begin{equation}
\label{3.1}
\frac{\partial f_h}{\partial t}-\gamma \frac{\partial}{\partial
{\bf v}}\cdot {\bf v} f_h=J[\mathbf{v}|f_h,f_h].
\end{equation}
The balance equations \eqref{2.7}--\eqref{2.9} yield $\partial_t n=0$, $\partial_t \mathbf{U}=\mathbf{0}$ and
\beq
\label{3.3}
\partial_t T=-(\zeta+2\gamma) T.
\eeq
Upon deriving Eq.\ \eqref{3.3} we have accounted for that the heat flux vanishes and the pressure tensor is diagonal, namely, $P_{ij}=p\delta_{ij}$. In the case of model A ($q=0$), $\nu_0$ does not depend on time and the solution to Eq.\ \eqref{3.3} is simply
\beq
\label{3.3.1}
\frac{T(t)}{T_0}=e^{-(2\gamma+\zeta)t},
\eeq
where $T_0$ is the initial temperature. On the other hand, in the case of model B with $q=\frac{1}{2}$, $\nu_0(t) \propto \sqrt{T(t)}$ and the solution to Eq.\ \eqref{3.3} for a three-dimensional system can be cast into the form \cite{YZHH13}
\beq
\label{3.3.2}
\frac{T(t)}{T_0}=\frac{4\gamma_0^{*2}e^{-2\gamma_0^* t^*}}{\left[2\gamma_0^*+\zeta^*\left(1-e^{-\gamma_0^* t^*}\right)\right]^2},
\eeq
where $\gamma_0^*\equiv \gamma/\nu_0(T_0)$, $\zeta^*\equiv \zeta/\nu_0=((d+2)/4d)(1-\al^2)$ and $t^*\equiv \nu_0(T_0) t$. To illustrate the time dependence of the temperature, Fig.\ \ref{fig0bis} shows the ratio $T(t)/T_0$ versus the (dimensionless) time $t^*$ for models A and B (with $q=\frac{1}{2}$) for the (initial) reduced friction coefficient $\gamma_0^*=0.1$ and the coefficient of restitution $\al=0.8$. The dry granular limit case ($\gamma_0^*=0$) of model B is also presented for comparison. As expected, the temperature decays in time more slowly in the dry limit case than in the case of viscous suspensions. Moreover, we observe that this decay is more pronounced in the case of model A (where the collision frequency $\nu_0$ is constant) than in the case of model B.
\begin{figure}
{\includegraphics[width=0.4\columnwidth]{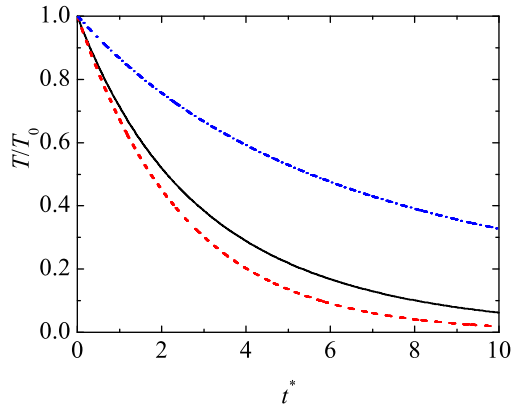}}
\caption{(Color online) Temperature versus (dimensionless) time for a three dimensional system with $\gamma_0^*=0.1$ and $\al=0.8$. The solid line corresponds to model B with $q=\frac{1}{2}$, the dashed red line corresponds to model A and the blue dash-dotted line corresponds to the results of model B in the dry granular case ($\gamma_0^*=0$).
\label{fig0bis}}
\end{figure}

In the hydrodynamic regime, since the time dependence of $f_h$ only occurs through the granular temperature $T$, then
\beq
\label{3.4}
\frac{\partial f_h}{\partial t}=
\frac{\partial f_h}{\partial T}\frac{\partial T}{\partial t}=-(\zeta+2\gamma)T\frac{\partial f_h}{\partial T},
\eeq
and the Boltzmann equation \eqref{3.1} becomes
\beq
\label{3.5}
-(\zeta+2\gamma)T\frac{\partial f_h}{\partial T}-\gamma \frac{\partial}{\partial
{\bf v}}\cdot {\bf v} f_h=J[\mathbf{v}|f_h,f_h].
\eeq

In the absence of the viscous drag force ($\gamma=0$), Eq.\ \eqref{3.5} admits the solution \cite{NE98}
\beq
\label{3.6}
f_h(\mathbf{v})=n v_0^{-d} \varphi_h (\mathbf{c}),
\eeq
where the scaling distribution $\varphi_h$ is an unknown function of the dimensionless velocity
\beq
\label{3.7}
\mathbf{c}=\frac{\mathbf{v}}{v_0},
\eeq
where we recall that $v_0\equiv \sqrt{2T/m}$ is the thermal velocity. When $\gamma \neq 0$, according to the previous results \cite{MGV14,MVG13,GMT12} derived for the complete model of suspensions (drag force plus stochastic force), the scaled distribution $\varphi_h$ could have an additional dependence on the granular temperature through the dimensionless friction coefficient $\gamma^*\equiv \gamma/\nu_0$. On the other hand, it can be seen by direct substitution that the form \eqref{3.6} is also a solution of Eq.\ \eqref{3.5} and hence $\varphi_h$ does not explicitly depend on $\gamma^*$. Thus,
\beq
\label{3.7.1}
T\frac{\partial f_h}{\partial T}=-\frac{1}{2}\frac{\partial}{\partial
{\bf v}}\cdot {\bf v} f_h,
\eeq
and Eq.\ \eqref{3.5} reduces to
\beq
\label{3.7.2}
\frac{1}{2}\zeta \frac{\partial}{\partial{\bf v}}\cdot {\bf v} f_h=J[f_h,f_h].
\eeq
Equation \eqref{3.7.2} is fully equivalent to the one obtained in the HCS of a dry granular gas (namely, when $\gamma^*=0$).

To confirm the scaling \eqref{3.6}, let us first analyze the evolution of the kurtosis or fourth-cumulant
\begin{equation}
\label{3.10}
a_{2}=\frac{1}{d(d+2)}\frac{m^2}{nT^2}\int\; \dd{\bf v}\; v^4 f_h(\mathbf{v})-1.
\end{equation}
Although the exact form of the homogeneous distribution function is not known, the knowledge of $a_2$ provides an indirect information of the deviation of $\varphi_h$ from its Gaussian form. In order to determine $a_2(t)$, we multiply Eq.\ \eqref{3.1} by $v^4$ and integrate over velocity. The result can be written as
\beq
\label{3.13}
\frac{\partial a_2}{\partial \tau}+\omega_{4|0}^*a_2=
\frac{d}{d+2}\left(\lambda_1^*-\frac{d+2}{d}\omega_{4|0}^*\right),
\eeq
where
\beq
\label{3.13.1}
\omega_{4|0}^*\equiv \frac{\nu_{4|0}-2\zeta}{\nu_0}=\frac{(1+\al)^2(4d-7+6\al-3\al^2)}{16 d},
\eeq
$\lambda_1^*\equiv \lambda_1/\nu_0$, and
\beq
\label{3.13.2}
\tau(t)=\int_0^t\; dt' \nu_0(t').
\eeq
The dimensionless time scale $\tau$ is therefore an average number of
collisions per particle in the time interval between $0$ and $t$. The solution to Eq.\ \eqref{3.13} is
\beq
\label{3.13.2bis}
a_2(\tau)=a_2(0)e^{-\omega_{4|0}^* \tau}+a_{2,\text{dry}},
\eeq
where $a_2(0)$ denotes the initial value of $a_2$ and
\beq
\label{3.13.3}
a_{2,\text{dry}}=\frac{d}{d+2}\frac{\lambda_1^*-\frac{d+2}{d}\omega_{4|0}^*}
{\omega_{4|0}^*}=\frac{6(1-\al^2)^2}{4d-7+3\al(2-\al)}
\eeq
is the value of $a_2$ in the case of a \emph{dry} granular gas \cite{S03}. For long times (hydrodynamic solution), since $\omega_{4|0}^*>0$ then $a_2\to a_{2,\text{dry}}$ and the results obtained for the (scaled) fourth-degree moment of $f_h$ in the presence or in the absence of the drag force are the same.

A similar analysis to the one carried out for $a_2$ can be made for the remaining (isotropic) moments
\beq
\label{3.14}
M_{2k}=\int\; \dd\mathbf{v} \; v^{2k} f_h=n \left(\frac{2T}{m}\right)^{k}\int\; \dd\mathbf{c} \; c^{2k} \varphi_h\equiv n \left(\frac{2T}{m}\right)^{k} M_{2k}^*,
\eeq
where the second identity defines the (dimensionless) moments $M_{2k}^*$ of degree $2k$. We want to see if actually the hydrodynamic expressions of $M_{2k}^*$ are identical to those obtained for a dry granular gas. To get those moments, we multiply both sides of Eq.\ \eqref{3.1} by $v^{2k}$ and integrate over velocity. After some algebra, we achieve the result
\beq
\label{3.15}
\frac{\partial M_{2k}^*}{\partial \tau}+\omega_{2k|0}^*M_{2k}^*=\sum_{k',k''}^\dagger \lambda_{k'k''}^*M_{2k'}^*M_{2k''}^*,
\eeq
where  $\omega_{2k|0}^*=\nu_{2k|0}^*-k\zeta^*$ and the dagger in the summation denotes the constraint $k'+k''<k$. The dimensionless quantities $\nu_{2k|0}^*$ and $\lambda_{k'k''}^*$ are nonlinear functions of the coefficient of restitution $\al$ but they are independent of the drag coefficient $\gamma^*$. In addition, upon deriving Eq.\ \eqref{3.15} use has been made of the mathematical structure of the collision operator for IMM that implies that a collisional moment of degree $2k$ can be expressed in terms of velocity moments of a degree less than or equal to $2k$. Assuming that the velocity moments $M_{2k'}$ of degree $2k'$ smaller than $2k$ have reached their \emph{steady} (dry) values (independent of the initial conditions), the solution of \eqref{3.15} can be cast into the form
\beq
\label{3.16}
M_{2k}^*(\tau)=M_{2k}^*(0)e^{-\omega_{2k|0}^* \tau}+M_{2k,\text{dry}}^*,
\eeq
where
\beq
\label{3.17}
M_{2k,\text{dry}}^*=-\omega_{2k|0}^{*-1}\sum_{k',k''}^\dagger \lambda_{k'k''}^*M_{2k',\text{dry}}^*M_{2k''\text{dry}}^*.
\eeq
Thus, for long times, if $\omega_{2k|0}^{*}>0$ then $M_{2k}^*(\tau)\to M_{2k,\text{dry}}^*$ and hence, the hydrodynamic expression of the (reduced) velocity moments $M_{2k}^*$ is fully consistent with the scaling solution \eqref{3.6} since they do not have an explicit dependence on $\gamma^*$.

\section{Chapman-Enskog method}
\label{sec4}

Let us assume that we slightly disturb the homogeneous time-dependent state studied in section \ref{sec3} by small spatial perturbations. In this case, the momentum and heat fluxes are not zero and the corresponding transport coefficients can be identified. The evaluation of these coefficients as functions of both the coefficient of restitution $\al$ and the friction coefficient $\gamma$ is the main goal of the present work.

Since the strength of the spatial gradients is small, the Boltzmann equation \eqref{5.3} is solved by means of the Chapman-Enskog method \cite{CC70} conveniently adapted for inelastic collisions. The Chapman-Enskog method assumes the existence of a \emph{normal} solution in which all the space and time dependence of the distribution function only occurs through a functional dependence on the hydrodynamic fields, i.e.,
\begin{equation}
f({\bf r},{\bf v},t)=f\left[{\bf v}|n ({\bf r}, t), {\bf U}({\bf r}, t), T({\bf r}, t) \right] \;.
\label{4.1}
\end{equation}
The notation on the right hand side indicates a functional dependence on the density, temperature and flow velocity. This functional dependence can be made local by an expansion of $f({\bf r},{\bf v},t)$ in powers of the spatial gradients of $n$, $\mathbf{U}$, and $T$:
\begin{equation}
f=f^{(0)}+f^{(1)}+f^{(2)}+\cdots \;, \label{4.2}
\end{equation}
where the approximation $f^{(k)}$ is of order $k$ in spatial gradients. In addition, to collect the different level of approximations in Eq.\ \eqref{5.3}, one has to characterize the magnitude of the drag coefficient $\gamma$ and the velocity difference $\Delta \mathbf{U}$ relative to the gradients as well. As in recent previous studies on suspensions \cite{GTSH12}, the parameter $\gamma$ is taken to be at least of zeroth-order in gradients. Another different consideration is given to $\Delta \mathbf{U}$ since $\mathbf{U}$ relaxes towards $\mathbf{U}_g$ after a transient period in the absence of spatial gradients [see Eq.\ \eqref{2.8} at zeroth-order]. In this case, the term $\Delta \mathbf{U}$ must be considered to be at least of first order in gradients.

The expansion \eqref{4.2} yields the corresponding expansions for the fluxes when one substitutes \eqref{4.2} into their definitions \eqref{2.10} and \eqref{2.11}:
\beq
\label{4.3}
{\sf P}={\sf P}^{(0)}+{\sf P}^{(1)}+\ldots, \quad \mathbf{q}=\mathbf{q}^{(0)}+\mathbf{q}^{(1)}+\ldots.
\eeq
In contrast to the results for IHS \cite{NE98}, the cooling rate of IMM is exactly given by the expression \eqref{2.22} and so, $\zeta^{(k)}=0$ for $k\geq 1$.  Finally, as usual in the Chapman-Enskog method, the time derivative is also expanded as
\beq
\label{4.4}
\partial_t=\partial_t^{(0)}+\partial_t^{(1)}+\ldots,
\eeq
where the action of each operator $\partial_t^{(k)}$ is obtained from the macroscopic balance equations \eqref{2.7}--\eqref{2.9} when one represents the fluxes and the cooling rate in their corresponding series expansion \eqref{4.3}. In this paper, we will restrict our calculations to the Navier-Stokes hydrodynamic order (first order contributions to the fluxes). The Burnett hydrodynamic equations (second order contributions to the fluxes) of a dry granular gas of IMM have been recently derived in Ref.\ \cite{NGS14}.

\subsection{Zeroth-order approximation}

To zeroth-order, the Boltzmann equation \eqref{5.3} for $f^{(0)}$ reads
\begin{equation}
\label{4.5}
\partial_t^{(0)}f^{(0)}-\gamma \frac{\partial}{\partial {\bf V}}\cdot {\bf V} f^{(0)}=J[f^{(0)},f^{(0)}].
\end{equation}
The balance equations at zeroth-order give $\partial_t^{(0)}n=\partial_t^{(0)} U_i=0$ and
\beq
\label{4.5.1}
\partial_t^{(0)} T=-(\zeta+2\gamma)T.
\eeq
Since $f^{(0)}$ qualifies as a normal solution, then
\beq
\label{4.5.2}
\partial_t^{(0)}f^{(0)}=\frac{\partial f^{(0)}}{\partial n}\partial_t^{(0)}n+\frac{\partial f^{(0)}}{\partial U_i}\partial_t^{(0)}U_i+
\frac{\partial f^{(0)}}{\partial T}\partial_t^{(0)}T=\frac{1}{2}(\zeta+2\gamma)\frac{\partial}{\partial {\bf V}}\cdot {\bf V} f^{(0)},
\eeq
where in the last step we have taken into account that $f^{(0)}$ depends on $\mathbf{U}$ through its dependence on $\mathbf{V}$. Substitution of Eq.\ \eqref{4.5.2} into Eq.\ \eqref{4.5} yields
\beq
\label{4.6}
\frac{1}{2}\zeta \frac{\partial}{\partial {\bf V}}\cdot {\bf V} f^{(0)}=J[f^{(0)},f^{(0)}].
\eeq
A solution to Eq.\ \eqref{4.6} is given by the local version of the time-dependent scaled distribution \eqref{3.6}. The isotropic properties of $f^{(0)}$ lead to $P_{ij}^{(0)}=p\delta_{ij}$ and $\mathbf{q}^{(0)}=\mathbf{0}$.

\section{First-order approximation: Navier-Stokes transport coefficients}
\label{sec5}

The analysis to first order in the gradients is worked out in the Appendix \ref{appA}. The first order velocity distribution function $f^{(1)}(\mathbf{V})$ verifies the kinetic equation
\beq
\label{5.5}
\left(\partial_{t}^{(0)}+{\cal L}\right)f^{(1)}-\gamma \frac{\partial}{\partial
{\bf V}}\cdot {\bf V} f^{(1)}=\mathbf{A}\cdot \nabla \ln T+\mathbf{B}\cdot \nabla \ln n +
C_{ij}\frac{1}{2}\left(\nabla_i U_j+\nabla_j U_i-\frac{2}{d}\delta_{ij}\nabla\cdot \mathbf{U}\right),
\eeq
where
\begin{equation}
\label{5.2}
{\cal L}f^{(1)}=-\left(J[f^{(0)},f^{(1)}]+J[f^{(1)},f^{(0)}]\right)
\end{equation}
is the linearized Boltzmann collision operator and the quantities $\mathbf{A}(\mathbf{V})$, $\mathbf{B}(\mathbf{V})$, and $C_{ij}(\mathbf{V})$ are given by Eqs.\ \eqref{a5}--\eqref{a7}, respectively. It must noted that for $q=\frac{1}{2}$, Eq.\ \eqref{5.5} has the same structure as that of the Boltzmann equation for IHS \cite{GMV13}. The only difference between IMM and IHS lies in the explicit form of the operator ${\cal L}$ that prevents to achieve exact results in the case of IHS.

Although the first-order distribution $f^{(1)}(\mathbf{V})$ is not explicitly known for IMM, the fact that the collisional moments of ${\cal L}f^{(1)}$ can be exactly computed opens the possibility of determining the Navier-Stokes transport coefficients. They are defined through the constitutive equations
\beq
\label{5.11}
P_{ij}^{(1)}=-\eta\left( \nabla_{i}U_{j}+\nabla_{j}U_{i}-\frac{2}{d}\delta _{ij}\nabla \cdot
\mathbf{U} \right),
\eeq
\beq
\label{5.15}
\mathbf{q}^{(1)}=-\kappa \nabla T-\mu \nabla n,
\eeq
where $\eta$ is the shear viscosity coefficient, $\kappa$ is the thermal conductivity coefficient and $\mu$ is a new transport coefficient not present for ordinary gases. The evaluation of those transport coefficients will be carried out in this section. Let us consider each flux separately.

\subsection{Pressure tensor}

The first-order contribution to the pressure tensor is
\beq
\label{5.8}
{\sf P}^{(1)}=\int\; \dd \mathbf{v}\; m \mathbf{V} \mathbf{V} f^{(1)}(\mathbf{V}).
\eeq
In order to determine $P_{ij}^{(1)}$, we multiply both sides of Eq.\ \eqref{5.5} by $m V_i V_j$ and integrates over velocity. After some algebra, one gets
\beq
\label{5.9}
\left(\partial_t^{(0)}+\nu_{0|2} \right)P_{ij}^{(1)}+2\gamma P_{ij}^{(1)}=-p \left(\nabla_i U_j+
\nabla_j U_i-\frac{2}{d}\delta_{ij}\nabla \cdot \mathbf{U}\right),
\eeq
where use has been made of Eq.\ \eqref{2.13} to first-order, namely,
\beq
\label{5.10}
\int\; \dd \mathbf{v}\; m V_i V_j\; {\cal L}f^{(1)}=\nu_{0|2}P_{ij}^{(1)},
\eeq
where $\nu_{0|2}$ is defined by Eq.\ \eqref{2.16}. The solution to Eq.\ \eqref{5.9} is given by Eq.\ \eqref{5.11} where the shear viscosity $\eta$ verifies the time-dependent equation
\beq
\label{5.12}
\left(\partial_t^{(0)}+\nu_{0|2} \right)\eta+2\gamma \eta =p.
\eeq
In the hydrodynamic regime, it is expected that the shear viscosity coefficient $\eta$ can be written as
\beq
\label{5.12.1}
\eta=\eta_0 \eta^*(\al,\gamma^*), \quad \gamma^*(T)\equiv \gamma/\nu_0(T),
\eeq
where $\eta_0=p/\nu_0$ is the Navier-Stokes shear viscosity of a dilute elastic gas in the absence of the drag force. The dimensionless function $\eta^*$ can depend on temperature through its dependence on the (reduced) friction coefficient $\gamma^*$. Since $\eta_0 \propto T^{1-q}$ and $\gamma^*\propto T^{-q}$, then
\beq
\label{5.13}
\partial_t^{(0)}\eta =\eta^*\partial_t^{(0)}\eta_0+\eta_0 \partial_t^{(0)}\eta^*=
\left[\eta^*(\partial_T \eta_0)+\eta_0 (\partial_T \eta^*)\right](\partial_t^{(0)} T)
=-(\zeta+2\gamma)\eta_0\left[(1-q)\eta^*-q\gamma^*\frac{\partial \eta^*}{\partial \gamma^*}\right].
\eeq
Consequently, in dimensionless form, Eq.\ \eqref{5.12} can be written as
\beq
\label{5.13.1}
-\left(\zeta^*+2\gamma^*\right)\left[(1-q)\eta^*-q\gamma^*\frac{\partial \eta^*}{\partial \gamma^*}\right]+(\nu_{0|2}^*+2\gamma^*)\eta^*=1,
\eeq
where $\nu_{0|2}^*\equiv \nu_{0|2}/\nu_0$.

\begin{figure}
{\includegraphics[width=0.4\columnwidth]{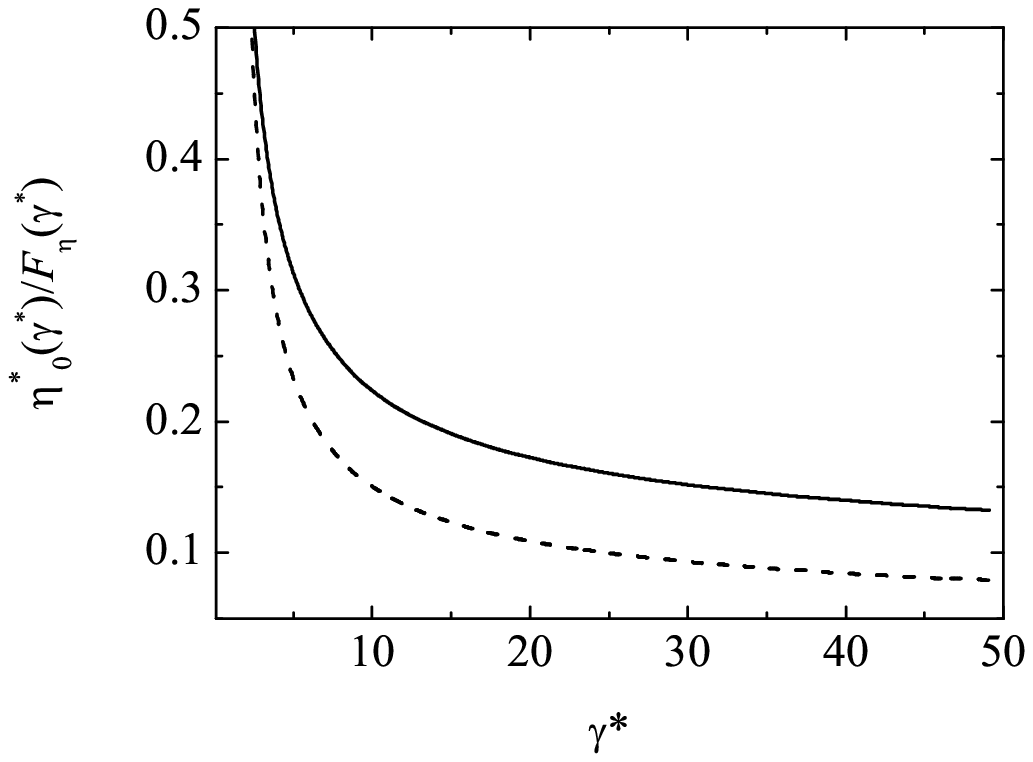}}
\caption{Plot of the ratio $\eta_0^*(\gamma^*)/F_\eta(\gamma^*)$ versus the dimensionless friction coefficient $\gamma^*$ for $d=3$, $\al=0.5$ and two different values of the interaction parameter $q$: $q=\frac{1}{2}$ (solid line) and $q=\frac{1}{4}$ (dashed line).
\label{fig0}}
\end{figure}

\begin{figure}
{\includegraphics[width=0.4\columnwidth]{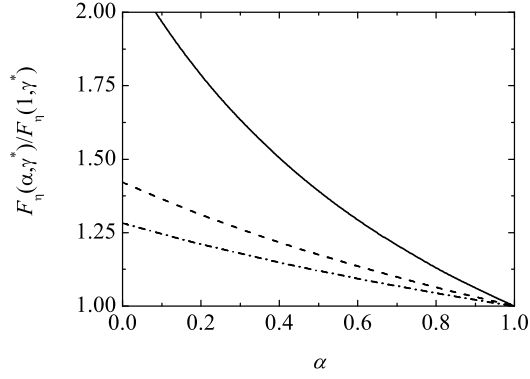}}
\caption{Plot of the ratio $F_\eta(\al, \gamma^*)/F_\eta(1, \gamma^*)$ as a function of the coefficient of restitution $\al$ for Model B with $q=\frac{1}{2}$ and three different values of $\gamma^*$: $\gamma^*=0$ (solid line), $\gamma^*=10$ (dotted line), and $\gamma^*=50$ (dash-dotted line).
\label{fig1}}
\end{figure}
In the case of a \emph{dry} granular gas ($\gamma^*=0$), the solution to Eq.\ \eqref{5.13.1} is
\beq
\label{5.13.2}
\eta_\text{dry}^*=\frac{1}{\omega_{0|2}^*+q\zeta^*},
\eeq
where
\beq
\label{5.14.3}
\omega_{0|2}^*\equiv \nu_{0|2}^*-\zeta^*=\frac{(1+\al)^2}{2}.
\eeq
Equation \eqref{5.13.2} is consistent with previous results derived for IMM when $q=\frac{1}{2}$ \cite{S03}. In the case of Model A ($q=0$), $\gamma^*\equiv \text{const.}$ and Eq.\ \eqref{5.13.1} becomes a simple algebraic equation independent of $\gamma^*$ whose solution is
\beq
\label{5.14.1}
\eta^*=\frac{1}{\omega_{0|2}^*}.
\eeq
Thus, for Model A, the (reduced) shear viscosity $\eta^*$ does not explicitly depend on the friction parameter and so, the drag force plays a neutral role on the momentum transport for this simple interaction model. This behavior is also present in the case of ordinary (elastic) Maxwell gases under uniform shear flow \cite{DSBR86} where there is a close relationship between the distribution functions with and without the drag force \eqref{5.0} (with $\mathbf{U}_g=\mathbf{U}$). However, such a relationship does not exist when other interaction potentials for ordinary gases are considered \cite{GS03}.

The case of Model B ($q\neq 0$) is more intricate since the (reduced) friction coefficient is also a function of time ($\gamma^*(t)\propto T(t)^{-q}$). In this case, the general solution to Eq.\ \eqref{5.13.1} can be written as
\beq
\label{5.14.1bis}
\eta^*(\al,\gamma^*)=C \eta_0^*(\al,\gamma^*) +F_\eta(\al,\gamma^*),
\eeq
where $C$ is a constant to be determined from the initial conditions and
\beq
\label{5.14.2bis}
\eta_0^*(\al,\gamma^*)=\exp\left\{\frac{1}{q\zeta^*}\left[\omega_{0|2}^*
\ln(2\gamma^*+\zeta^*)
-(\omega_{0|2}^*+q\zeta^*)\ln(2\gamma^*)\right]\right\},
\eeq
\beqa
\label{5.14.2}
F_\eta(\al,\gamma^*)&=&\frac{\eta_\text{dry}^*}{
\zeta^*\left(\omega_{0|2}^*+2q \zeta^*\right)}\left(1+\frac{2\gamma^*}{\zeta^*}\right)^{\omega_{0|2}^*/q\zeta^*}
\zeta^*\left(\omega_{0|2}^*+2q \zeta^*\right)\; {_2F_1}\left(\frac{\omega_{0|2}^*}{q\zeta^*}, 1+\frac{\omega_{0|2}^*}{q\zeta^*},2+\frac{\omega_{0|2}^*}{q\zeta^*},
-\frac{2\gamma^*}{\zeta^*}\right)
\nonumber\\
& &
-2\gamma^*\left(\omega_{0|2}^*+q \zeta^*\right) \; {_2F_1}\left( 1+\frac{\omega_{0|2}^*}{q\zeta^*},2+\frac{\omega_{0|2}^*}{q\zeta^*}, 3+\frac{\omega_{0|2}^*}{q\zeta^*},-\frac{2\gamma^*}{\zeta^*}\right),
\eeqa
where $_2F_1\left(a,b;c;z\right)$ is the hypergeometric function \cite{AS72}. In the absence of the drag force ($\gamma^*=0$), as expected $F_\eta(\al,0)=\eta_\text{dry}^*$.

A hydrodynamic expression for the shear viscosity, independent of the initial conditions, is expected to hold after a transient period. In the long-time limit,  $T(t)\to 0$ and so, $\gamma^*\to \infty$. Thus, to analyze whether the system reaches a hydrodynamic regime we have to see if actually the ratio $\eta_0^*/F_\eta$ goes to zero when $\gamma^*\to \infty$. Although we have not shown it analytically, we have numerically observed this behavior for different values of $\al$ and $q$. As an illustration, Fig.\ \ref{fig0} shows $\eta_0^*/F_\eta$ versus $\gamma^*$ for $\al=0.5$ and two different values of the interaction parameter $q$. It is quite apparent that $\lim_{\gamma^*\to \infty}\eta_0^*/F_\eta=0$ and hence, for sufficiently long times one can neglect the initial term in Eq.\ \eqref{5.14.1bis} and the hydrodynamic form of the shear viscosity coefficient $\eta$ for Model B is
\beq
\label{5.14.4}
\eta(\al,\gamma^*)=\eta_0 F_\eta(\al,\gamma^*).
\eeq
The simple expression \eqref{5.14.1} for Model A is recovered by taking the limit $q\to 0$ in Eq.\ \eqref{5.14.4}.

Figure \ref{fig1} shows the dependence of the ratio $F_\eta(\al, \gamma^*)/F_\eta(1, \gamma^*)$ on the coefficient of restitution $\al$ for Model B with $q=\frac{1}{2}$ and two different values of $\gamma^*$. We observe that the impact of the interstitial fluid on the shear viscosity increases with the collisional dissipation. Moreover, at a given value of $\al$, it is quite apparent that the magnitude of $\eta^*$ decreases as $\gamma^*$ increases and hence, the shear viscosity of a dry granular gas is larger than that of its corresponding gas-solid suspension.

\subsection{Heat flux vector}

The heat flux to first-order is
\beq
\label{5.16.0}
\mathbf{q}^{(1)}=\int\; \dd\mathbf{v}\; \frac{m}{2}V^2 \mathbf{V} f^{(1)}(\mathbf{V}).
\eeq
As in the case of the pressure tensor, to obtain $\mathbf{q}^{(1)}$ we multiply both sides of Eq.\ \eqref{5.5} by $\frac{m}{2} V^2 \mathbf{V}$ and integrate over $\mathbf{v}$. After some algebra, one gets
\beq
\label{5.16}
\partial_t^{(0)}\mathbf{q}^{(1)}+\left(\nu_{2|1}+3\gamma\right)\mathbf{q}^{(1)}=
-\frac{d+2}{2}\frac{p}{m}\left(1+2a_2\right)\nabla T -\frac{d+2}{2}\frac{T^2}{m}a_2\nabla n,
\eeq
where use has been made of Eq.\ \eqref{2.14} to first-order, namely,
\beq
\label{5.17}
\int \dd \mathbf{v}\; \frac{m}{2} V^2 \mathbf{V} {\cal L}f^{(1)}=\nu_{2|1} \mathbf{q}^{(1)},
\eeq
where $\nu_{2|1}$ is defined by Eq.\ \eqref{2.17}. In addition, upon writing Eq.\ \eqref{5.16}, the following results have been used:
\beq
\label{5.18}
\int\; \dd \mathbf{v}\; \frac{m}{2}V^2 V_i A_j(\mathbf{V})=
-\frac{d+2}{2}\frac{p T}{m}\delta_{ij}\left(1+2a_2\right),
\eeq
\beq
\label{5.19}
\int\; \dd \mathbf{v}\; \frac{m}{2}V^2 V_i B_j(\mathbf{V})=-\frac{d+2}{2}\frac{p T}{m}a_2\delta_{ij}.
\eeq

The solution to Eq.\ \eqref{5.16} is given by the constitutive equation \eqref{5.15}. As in the case of the shear viscosity, the coefficients $\kappa$ and $\mu$ appearing in Eq.\ \eqref{5.15} can be written as
\beq
\label{5.15.1bis}
\kappa=\kappa_0 \kappa^*(\al,\gamma^*), \quad \mu=\frac{T\kappa_0}{n}\mu^*(\al,\gamma^*),
\eeq
where
\beq
\label{5.15.1}
\kappa_0=\frac{d(d+2)}{2(d-1)}\frac{\eta_0}{m}
\eeq
is the Navier-Stokes thermal conductivity of a dilute elastic gas in the absence of the drag force. Note that the one-dimensional case ($d=1$) for $\kappa_0$ deserves some care since it diverges at $d=1$ \cite{NR02}. However, as we will show below the thermal conductivity $\kappa$ is well defined at $d = 1$ for dry granular gases ($\al < 1$).

The action of the operator $\partial_t^{(0)}$ over the heat flux $\mathbf{q}^{(1)}$ in Eq.\ \eqref{5.16} gives
\beqa
\label{5.15.2}
\partial_t^{(0)}\mathbf{q}^{(1)}&=&-(\partial_t^{(0)} \kappa)\nabla T-\kappa \nabla (\partial_t^{(0)}T)-(\partial_t^{(0)}\mu)\nabla n \nonumber\\
&=&\kappa_0\left\{2\left[\zeta+(2-q)\gamma\right]\kappa^*-q(\zeta+2\gamma)
\gamma^*\frac{\partial \kappa^*}
{\partial \gamma^*}\right\}\nabla T \nonumber\\
& +& \frac{T\kappa_0}{n}\left\{\zeta \kappa^*+(\zeta+2\gamma)\left[(2-q)\mu^*-q\gamma^*\frac{\partial \mu^*}
{\partial \gamma^*}\right]\right\}\nabla n.
\eeqa
The differential equations defining the transport coefficients $\kappa$ and $\mu$ can be obtained by substituting Eq.\ \eqref{5.15.2} into Eq.\ \eqref{5.16} and identifying coefficients of $\nabla T$ and $\nabla n$. In dimensionless form, the corresponding equations for $\kappa^*$ and $\mu^*$ are
\beq
\label{5.17}
\left[\omega_{2|1}^*-\frac{1}{2}\zeta^*-2\gamma^*\left(\frac{1}{2}-q\right)\right]\kappa^*
+q(\zeta^*+2\gamma^*)\gamma^*\frac{\partial \kappa^*}
{\partial \gamma^*}=\frac{d-1}{d}\left(1+2a_2\right),
\eeq
\beq
\label{5.18}
\left[\omega_{2|1}^*+\left(q-\frac{1}{2}\right)\left(\zeta^*+2\gamma^*\right)\right]\mu^*
+q(\zeta^*+2\gamma^*)\gamma^*\frac{\partial \mu^*}
{\partial \gamma^*}=\frac{d-1}{d}a_2+\zeta^*\kappa^*,
\eeq
where $\nu_{2|1}^*\equiv \nu_{2|1}/\nu_0$ and
\beq
\label{5.21}
\omega_{2|1}^*\equiv \nu_{2|1}^*-\frac{3}{2}\zeta^*=\frac{d-1}{4d}(1+\al)^2.
\eeq

In the absence of the gas phase ($\gamma^*=0$), the solution to Eqs.\ \eqref{5.17} and \eqref{5.18} is
\beq
\label{5.18.1}
\kappa_\text{dry}^*=\frac{d-1}{d}\frac{1+2a_2}{\omega_{2|1}^*-\frac{1}{2}\zeta^*},
\eeq
\beq
\label{5.18.2}
\mu_\text{dry}^*=\frac{\frac{d-1}{d}a_2+\zeta^*\kappa_\text{dry}^*}{\omega_{2|1}^*+
\left(q-\frac{1}{2}\right)\zeta^*}.
\eeq
When $q=\frac{1}{2}$, Eqs.\ \eqref{5.18.1} and \eqref{5.18.2} agree with those previously derived \cite{S03} for an undriven granular gas of IMM. As for the shear viscosity, the set of nonlinear differential equations \eqref{5.17} and \eqref{5.18} become a set of algebraic equations for Model A ($q=0$), whose solution is
\beq
\label{5.19}
\kappa^*= \frac{d-1}{d}\frac{1+2 a_2}{\omega_{2|1}^*-\frac{1}{2}\zeta^*-\gamma^*},
\eeq
\beq
\label{5.19.1}
\mu^*=\frac{\frac{d-1}{d}a_2+\zeta^*\kappa^*}{\omega_{2|1}^*-\frac{1}{2}\zeta^*-\gamma^*}.
\eeq
Equations \eqref{5.19} and \eqref{5.19.1} become unphysical when $\gamma^* \geq \omega_{2|1}^*-\frac{1}{2}\zeta^*$ since $\kappa^*$ and $\mu^*$ become divergent or negative. This singular behavior has been also found in the case of the self-diffusion coefficient of an ordinary Maxwell gas in the presence of a nonconservative drag force \cite{GSB90}.}

The solution to Eqs.\ \eqref{5.17} and \eqref{5.18} for Model B ($q\neq 0$) is much more intricate. On the other hand, an inspection to both equations shows that in the case $q=\frac{1}{2}$ a \emph{particular} (hydrodynamic) solution to them corresponds to the expressions of $\kappa^*$ and $\mu^*$ obtained in the dry limit case, namely, $\kappa^*=\kappa_\text{dry}^*$ (see Eq.\ \eqref{5.18.1}) and
\beq
\label{muihs}
\mu^*=\frac{\frac{d-1}{d}a_2+\zeta^*\kappa_\text{dry}^*}{\omega_{2|1}^*}
=\frac{d-1}{d}\frac{\zeta^*(1+2a_2)+a_2(\omega_{2|1}^*-
\frac{1}{2}\zeta^*)}{\omega_{2|1}^*(\omega_{2|1}^*-\frac{1}{2}\zeta^*)}.
\eeq
For general values of the interaction parameter $q$, the solution to Eq.\ \eqref{5.17} can be cast into the form
\beq
\label{5.20.1}
\kappa^*(\al,\gamma^*)=C \kappa_0^*(\al,\gamma^*)+F_\kappa(\al,\gamma^*),
\eeq
where $C$ is a constant to be determined from the initial conditions and
\beq
\label{5.20.1bis}
\kappa_0^*(\al,\gamma^*)=
\exp\left\{\frac{1}{q\zeta^*}\left[(\frac{1}{2}\zeta^*-\omega_{2|1}^*)
\ln(2\gamma^*)+(\omega_{2|1}^*-q\zeta^*)\ln(2\gamma^*+\zeta^*)\right]\right\},
\eeq
\beqa
\label{5.20}
F_\kappa(\al,\gamma^*)&=&\frac{\kappa_\text{dry}^*}{
\left(\zeta^*+2\gamma^*\right)
\left[\omega_{2|1}^*+(q-\frac{1}{2})\zeta^*\right]}
\left(1+\frac{2\gamma^*}{\zeta^*}\right)^{\omega_{2|1}^*/q\zeta^*}\nonumber\\
& & \times
\zeta^*\left[\omega_{2|1}^*+(q-\frac{1}{2})\zeta^*\right]\;
{_2F_1}\left(\frac{-\frac{1}{2q}+\omega_{2|1}^*}{q\zeta^*}, -1+\frac{\omega_{2|1}^*}{q\zeta^*},1-\frac{1}{2q}+\frac{\omega_{2|1}^*}{q\zeta^*},
-\frac{2\gamma^*}{\zeta^*}\right)
\nonumber\\
& &
+2\gamma^*\left(\frac{1}{2}\zeta^*-\omega_{2|1}^*\right) \;{_2F_1}\left(
1-\frac{1}{2q}+\frac{\omega_{2|1}^*}{q\zeta^*},\frac{\omega_{2|1}^*}{q\zeta^*},
2-\frac{1}{2q}+\frac{\omega_{2|1}^*}{q\zeta^*},
-\frac{2\gamma^*}{\zeta^*}\right).
\eeqa
It is important to note that the expression \eqref{5.20} for $F_\kappa$ (which is independent of the initial condition) is consistent with the particular solution \eqref{5.18.1} for Model B with $q=\frac{1}{2}$ and with Eq.\ \eqref{5.19} for Model A ($q=0$). Moreover, in the absence of the drag force ($\gamma^*=0$), as expected $F_\kappa(\al,0)=\kappa_\text{dry}^*$ and one recovers the results for dry granular gases.

\begin{figure}
{\includegraphics[width=0.4\columnwidth]{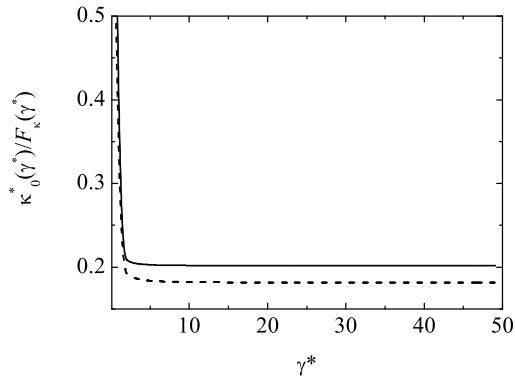}}
\caption{Plot of the ratio $\kappa_0^*(\gamma^*)/F_\kappa(\gamma^*)$ versus the dimensionless friction coefficient $\gamma^*$ for $d=3$, $\al=0.5$ and two different values of the interaction parameter $q$: $q=\frac{1}{4}$ (solid line) and $q=\frac{1}{3}$ (dashed line).
\label{fig2bis}}
\end{figure}

As for the shear viscosity, one expects that after a transient period, the coefficient $\kappa^*$ achieves its hydrodynamic value $F_\kappa$. To check it, we have to analyze the asymptotic behavior of the ratio $\kappa_0^*/F_\kappa$ in the long time limit ($\gamma^*\to \infty$). Figure \ref{fig2bis} shows $\kappa_0^*/F_\kappa$ versus $\gamma^*$ for $d=3$, $\al=0.5$ and two values of $q$. Although the function $\kappa_0^*/F_\kappa$ decreases as $\gamma^*$ increases, it decays much more slowly than in the case of the shear viscosity (see Fig.\ \ref{fig0}). In fact, for very large values of $\gamma^*$, the numerical results obtained for $\kappa_0^*/F_\kappa$ seem to indicate that this ratio reaches a constant asymptotic value (plateau) different from zero (for instance, for $q=\frac{1}{3}$ and $\al=0.5$, $\kappa_0^*/F_\kappa\simeq 0.1813$ when $\gamma^*\to \infty$). Therefore, for $q\leq \frac{1}{2}$, the presence of the drag force could prevent the existence of a hydrodynamic solution for $\kappa^*$ for the above range of values of the interaction parameter $q$. A similar behavior can be expected for the coefficient $\mu^*$ since the equation defining it (see Eq.\ \eqref{5.18}) involves to the thermal conductivity. The confirmation of the absence of hydrodynamic forms for the heat flux transport coefficients of model B could be achieved by numerically solving the Boltzmann equation by means of the direct simulation Monte Carlo (DSMC) method \cite{B94}. This is quite an interesting problem to be addressed in the near future.

\begin{figure}
{\includegraphics[width=0.4\columnwidth]{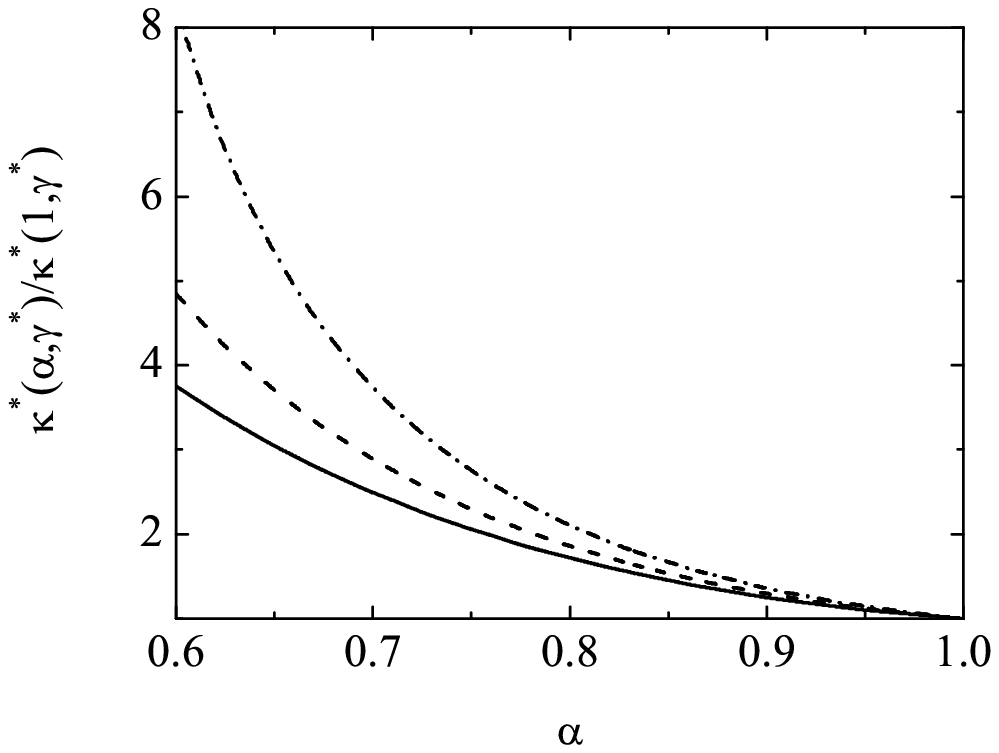}}
{\includegraphics[width=0.4\columnwidth]{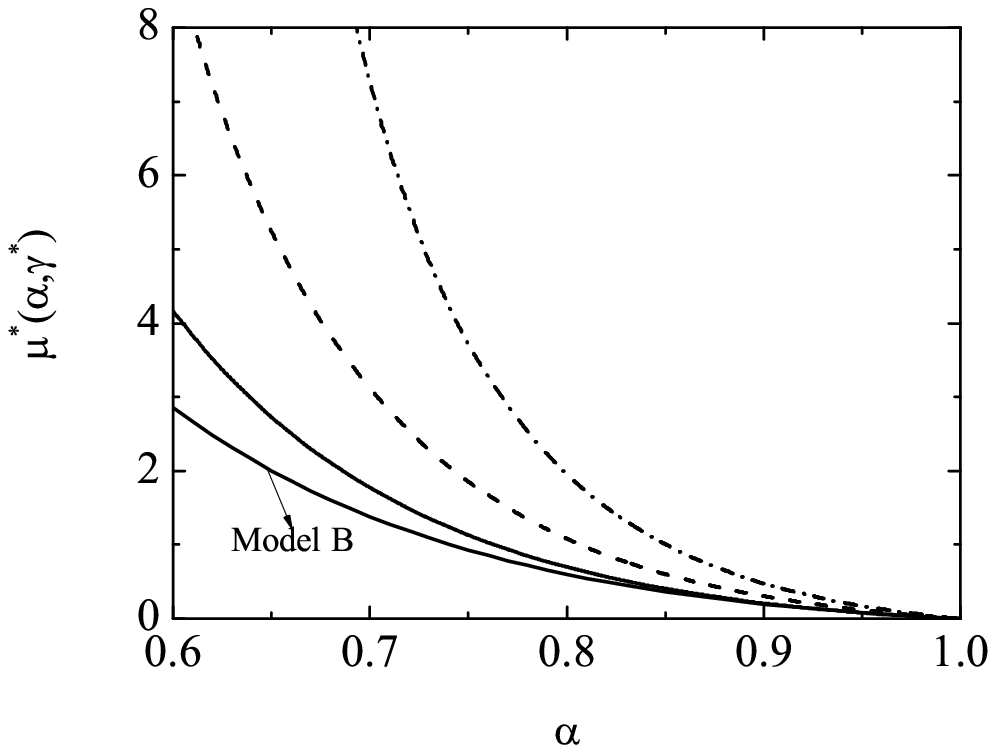}}
\caption{Plot of the ratio $\kappa^*(\al,\gamma^*)/\kappa^*(1,\gamma^*)$ (left panel) and $\mu^*(\al,\gamma^*)$ (right panel) as functions of the coefficient of restitution $\al$ for Model A ($q=0$) and for three different values of $\gamma^*$: $\gamma^*=0$ (solid lines), $\gamma^*=0.1$ (dashed lines) and $\gamma^*=0.2$ (dash-dotted lines). The results obtained for Model B with $q=\frac{1}{2}$ (which are independent of $\gamma^*$) have been also included. Note that in the case of the thermal conductivity the results of models A (with $\gamma^*=0$) and B (with $q=\frac{1}{2}$) are the same.
\label{fig2}}
\end{figure}

To illustrate the dependence of $\kappa^*$ and $\mu^*$ on both the coefficient of restitution $\al$ and the (reduced) friction coefficient $\gamma^*$, Fig.\ \ref{fig2} shows the ratio $\kappa^*(\al,\gamma^*)/\kappa^*(1,\gamma^*)$ and the dimensionless coefficient $\mu^*(\al,\gamma^*)$ as functions of $\al$ for three values of $\gamma^*$. We have considered here the most interesting physical models: Model A ($q=0$) and Model B with $q=\frac{1}{2}$. The first case corresponds to an interaction model closer to ordinary Maxwell molecules while the second case is closer to IHS. In the case of model B with $q=\frac{1}{2}$, we have plotted the particular (hydrodynamic) solutions given by Eq.\ \eqref{5.18.1} and for $\kappa^*$ and Eq.\ \eqref{muihs} for $\mu^*$. In addition, the (dimensionless) coefficient $\mu^*(\al,\gamma^*)$ is plotted rather than the ratio $\mu^*(\al,\gamma^*)/\mu^*(1,\gamma^*)$ since the latter diverges for elastic collisions ($\mu^*=0$ for $\al=1$). As in the case of the shear viscosity, the influence of the gas phase on the thermal conductivity becomes more significant as the dissipation increases. With respect to the coefficient $\mu^*$, at a given value of $\al$, we see that this coefficient decreases as the interaction becomes harder. Regarding the influence of the gas phase on $\mu^*$, we observe that the impact of $\gamma^*$ on $\mu^*$ is larger than the one predicted for the thermal conductivity.

\section{Some illustrative systems}
\label{sec6}

\subsection{Low mean-flow Reynolds numbers}

\begin{figure}
{\includegraphics[width=0.41\columnwidth]{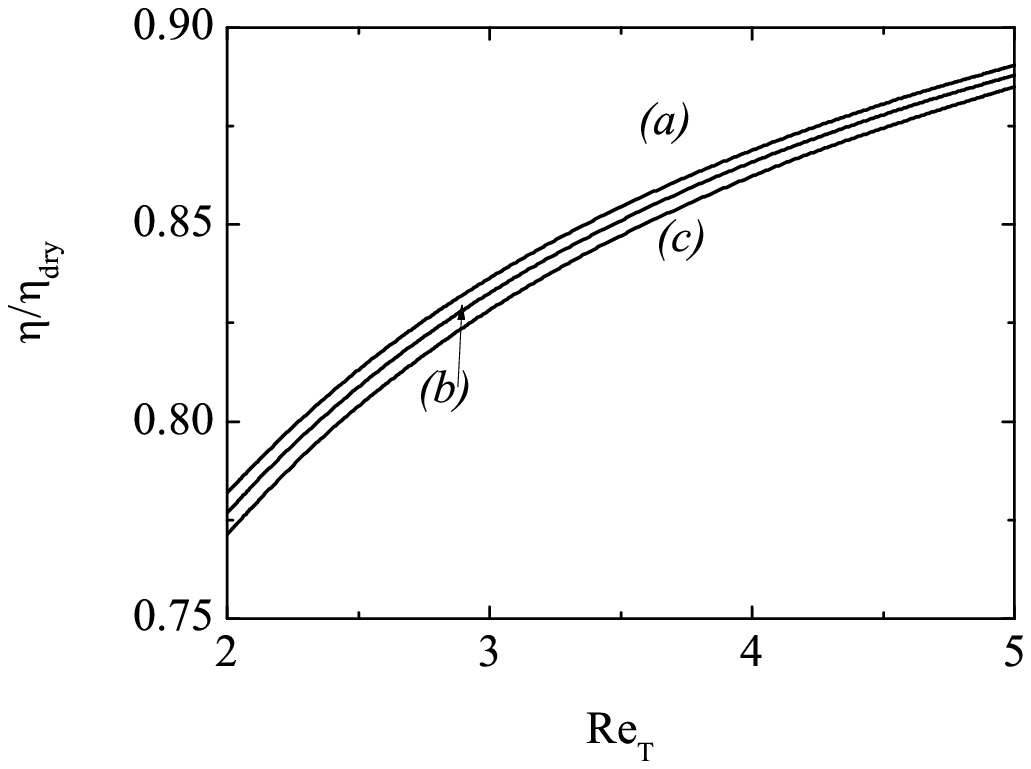}}
{\includegraphics[width=0.4\columnwidth]{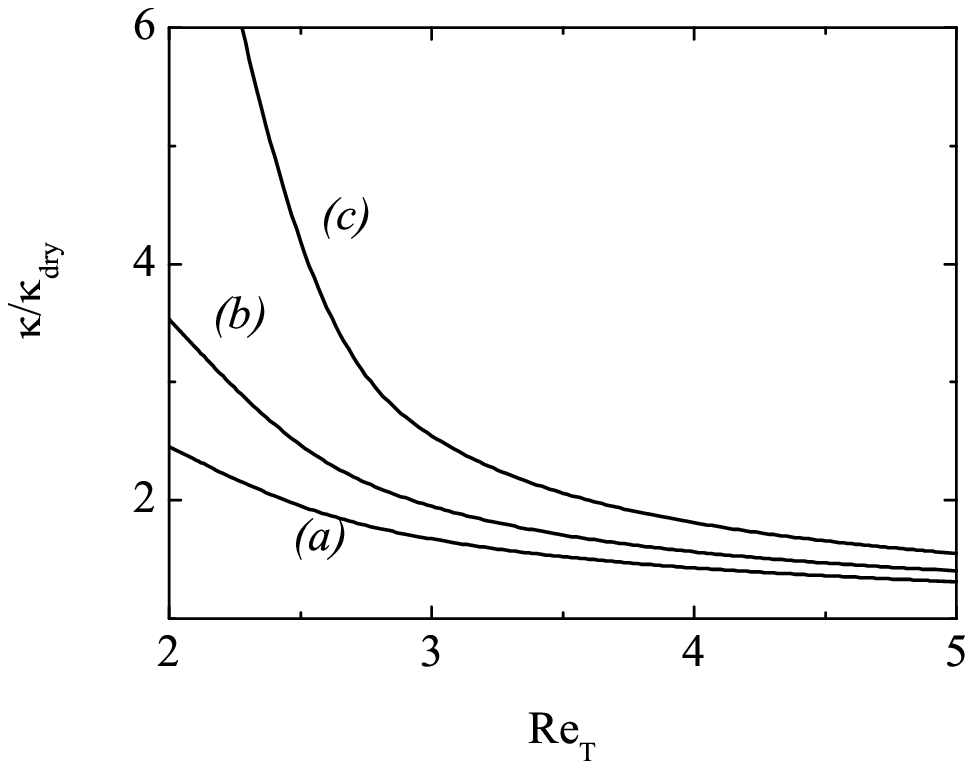}}
{\includegraphics[width=0.4\columnwidth]{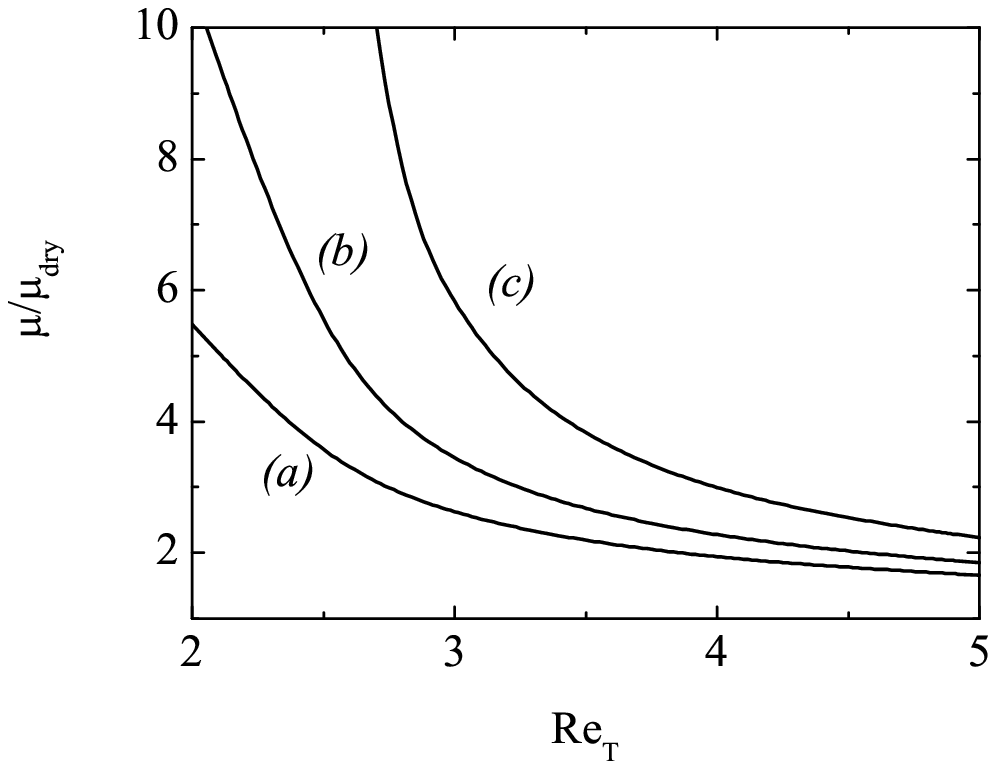}}
\caption{Plot of the ratios $\eta/\eta_\text{dry}$, $\kappa/\kappa_\text{dry}$ and $\mu/\mu_\text{dry}$ as functions of the Reynolds number $\text{Re}_T$ for $\phi=0.01$, $\rho_s/\rho_g=1000$ and three different values of the coefficient of restitution: $\al=0.9 (a)$, $\al=0.8 (b)$ and $\al=0.7 (c)$. In the case of the shear viscosity, $\eta$ is given by Eq.\ \eqref{5.14.2} with $q=\frac{1}{2}$ while the thermal conductivity $\kappa$ and the coefficient $\mu$ are given by Eqs.\ \eqref{5.19} and \eqref{5.19.1} of Model A ($q=0$), respectively.
\label{fig3}}
\end{figure}

Although the expressions derived in section \ref{sec5} for the Navier-Stokes transport coefficients of monodisperse gas-solid flows have been obtained in the framework of IMM, it is tempting to establish some connection with the results obtained for suspensions modeled as IHS. In particular, according to the results reported in Ref.\ \cite{GTSH12} for hard spheres ($d=3$), the (dimensionless) friction coefficient $\gamma^*$ can be written as
\beq
\label{6.1bis}
\gamma^*=\frac{3\pi}{\sqrt{2}\phi}\frac{\rho_g}{\rho_s} \text{Re}_\text{T}^{-1},
\eeq
where $\phi=(\pi/6)n\sigma^3$ is the solid volume fraction for spheres, $\sigma$ is the particle diameter, $\rho_g$ and $\rho_s$ are the mass density of gas and solid particles, respectively, and
\beq
\label{6.2bis}
\text{Re}_\text{T}=\frac{\rho_g \sigma}{\mu_g}\sqrt{\frac{T}{m}}
\eeq
is the Reynolds number associated with the particle velocity fluctuations. In Eq.\ \eqref{6.2bis}, $\mu_g$ is the dynamic viscosity of the gas phase. Note that the relation \eqref{6.1bis} only holds for low mean-flow Reynolds numbers and for very dilute systems \cite{GTSH12}. The dependence of the ratios $\eta/\eta_\text{dry}$, $\kappa/\kappa_\text{dry}$ and $\mu/\mu_\text{dry}$ on $\text{Re}_\text{T}$ is plotted in Fig.\ \ref{fig3} for $\phi=0.01$ (low-density granular system) with $\rho_s/\rho_g=1000$. Three different values of the coefficient of restitution are considered. Given that in the case of Model A, $\eta$ does not depend on the friction coefficient $\gamma$, we have plotted in Fig.\ \ref{fig3} the value of the shear viscosity defined by Eq.\ \eqref{5.14.2} for Model B with $q=\frac{1}{2}$. On the other hand, in the cases of $\kappa$ and $\mu$ we have plotted their corresponding \emph{simple} expressions \eqref{5.19} and \eqref{5.19.1}, respectively, derived for Model A ($q=0$). Moreover, in Fig.\ \ref{fig3} $\eta_\text{dry}$ is given by Eq.\ \eqref{5.13.2} with $q=\frac{1}{2}$ while $\kappa_\text{dry}$ and  $\mu_\text{dry}$ are given by Eqs.\ \eqref{5.18.1} and \eqref{5.18.2}, respectively, with $q=0$. It must be also remarked that the range of values of the Reynolds number as well as the values of $\phi$ and $\rho_s/\rho_g$ used in the above figure are typical values encountered in a circulating fluidized bed \cite{GTSH12}.

We observe first that the gas phase displays a larger impact on the shear viscosity for lower $\text{Re}_\text{T}$ while in the other extreme of higher $\text{Re}_\text{T}$, the granular limit ($\eta/\eta_\text{dry}\to 1$) is approached, as expected. It is also apparent that the gas phase effect on shear viscosity is more pronounced for higher dissipation levels (lower $\al$).
A comparison with the results derived in Ref.\ \cite{GTSH12} at the level of the shear viscosity of IHS shows a good qualitative agreement between both interaction models. Regarding the thermal conductivity, Fig.\ \ref{fig3} clearly shows a significant influence of the interstitial fluid on $\kappa$ for Model A, especially at lower Reynolds numbers. This contrasts with the results obtained here for $\kappa$ in the case of Model B with $q=\frac{1}{2}$ (which is the IMM closer to IHS) since the (exact) expression \eqref{5.20} for $\kappa$ turns out to be independent of $\gamma$ for this interaction model. On the other hand, this surprising result agrees qualitatively well with the findings of IHS \cite{GTSH12} since the latter shows a negligible impact of the gas phase on $\kappa$ over the range of parameters examined. Finally, in stark contrast with the shear viscosity, we see that the gas phase serves to increase the value of the coefficient $\mu$ (i.e., $\mu/\mu_\text{dry}>1$). This is consistent with the results of IHS \cite{GTSH12}. However, at a more quantitative level, the results for IHS are the opposite (see Figure 9 of Ref.\ \cite{GTSH12}) as those obtained here for IMM since the latter shows that the impact of gas phase on $\mu$ is more noticeable at higher dissipation levels (smaller $\al$).

\subsection{Steady states}

Apart from modeling the friction of solid particles with the surrounding fluid in gas-solid suspensions, the drag force \eqref{5.0} has been also used in nonequilibrium problems as a thermostatic force to achieve steady states. For instance, in the case of sheared ordinary fluids, the friction coefficient $\gamma$ is a (positive) shear-rate dependent function chosen to compensate for the viscous heating produced by shear work \cite{GS03,GSB90,EM90,MS00}. On the other hand, in the case of granular gases in homogenous states, when $\gamma<0$ the system is heated by an ``antidrag'' force chosen to exactly compensate for collisional cooling and reach a steady state. According to Eq.\ \eqref{3.3}, the condition $\partial_t T=0$ yields $\gamma=-\frac{1}{2}\zeta$, namely,
\beq
\label{6.3bis}
\gamma(\al)=-\frac{d+2}{8d}(1-\al^2)\nu_0.
\eeq
\begin{figure}
{\includegraphics[width=0.42\columnwidth]{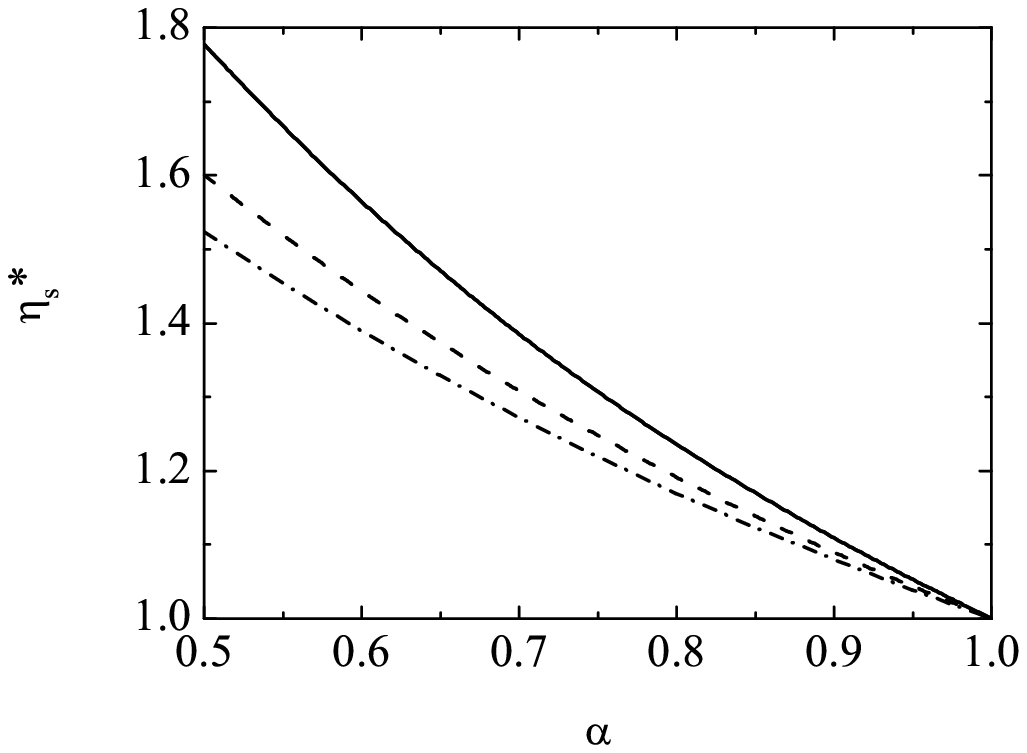}}
{\includegraphics[width=0.4\columnwidth]{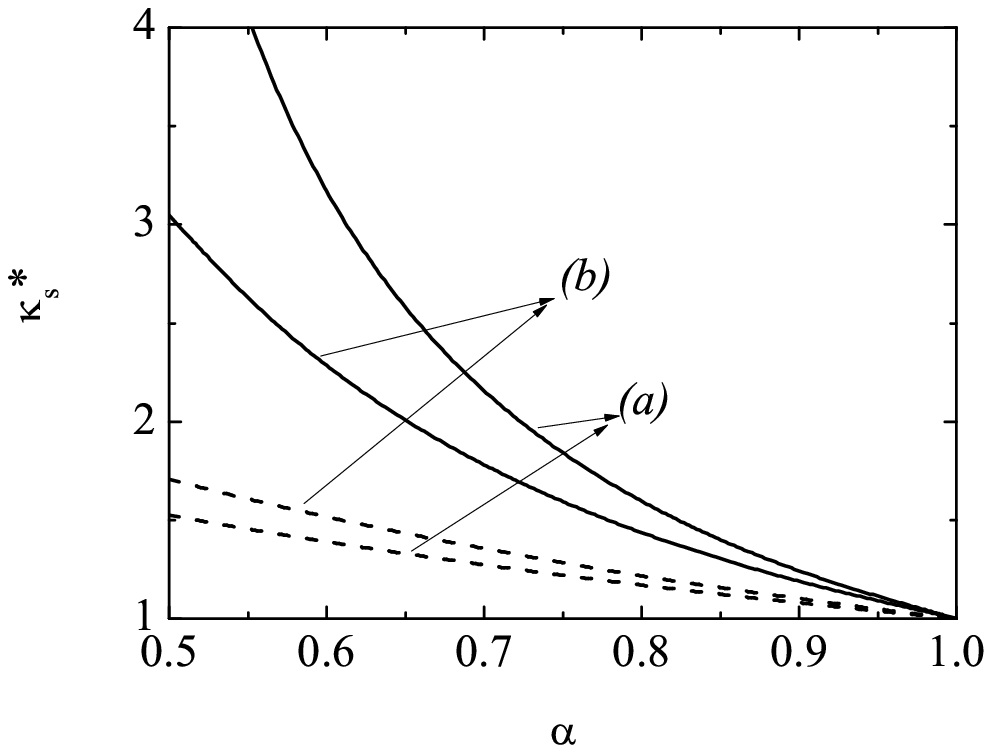}}
{\includegraphics[width=0.4\columnwidth]{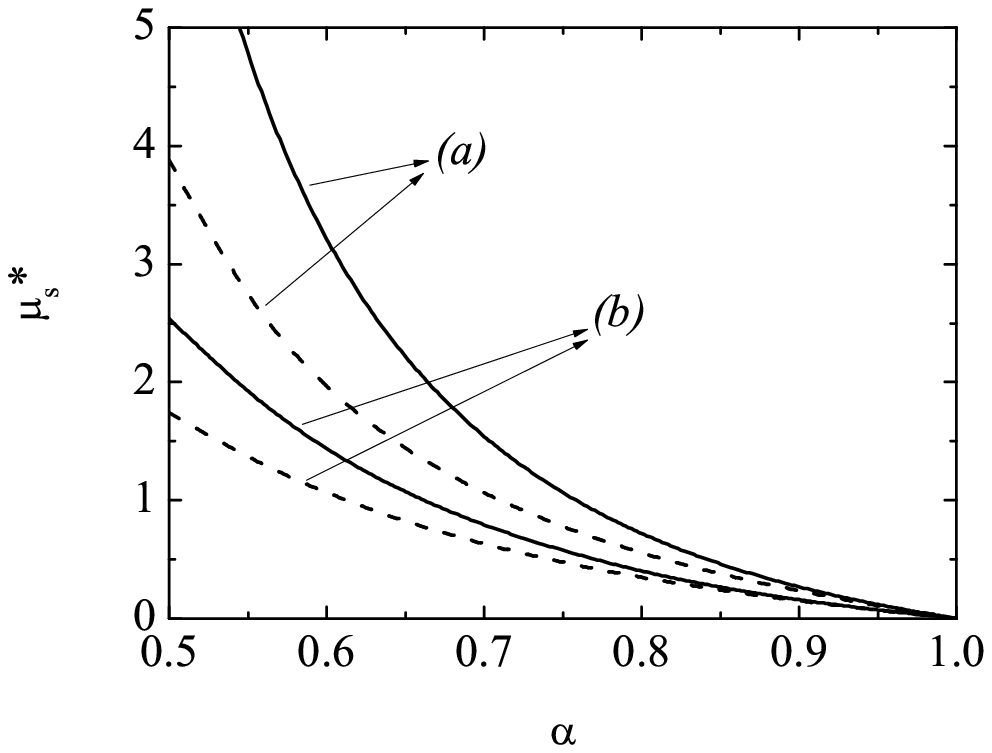}}
\caption{Plot of the \emph{steady} (dimensionless) transport coefficients $\eta_\text{s}^*$, $\kappa_\text{s}^*$ and $\mu_\text{s}^*$ as functions of the coefficient of restitution $\al$ as obtained from Model B with $q=\frac{1}{2}$ (solid lines) and from IHS (dashed lines) for disks $(a)$ and spheres $(b)$. In the case of the shear viscosity $\eta_\text{s}^*$, the solid line refers to Model B for both disks and spheres (its expression is independent of the dimensionality of the system) while the dash-dotted  and dashed lines correspond to $d=2$ and $d=3$, respectively, for IHS.
\label{fig4}}
\end{figure}
Therefore, $\gamma$ is taken as a negative coefficient coupled to the coefficient of restitution $\al$ \cite{GM02}. The expressions of the (reduced) transport coefficients $\eta_\text{s}^*$, $\kappa_\text{s}^*$ and $\mu_\text{s}^*$ in the steady state can be easily obtained from the general results derived in section \ref{sec5} by replacing $\gamma$ by its $\al$-dependent form \eqref{6.3bis}. They are given by
\beq
\label{6.4bis}
\eta_\text{s}^*=\frac{1}{\omega_{0|2}^*},
\eeq
\beq
\label{6.5bis}
\kappa_\text{s}^*=\frac{d-1}{d}\frac{1}{\omega_{2|1}^*-q\zeta^*},
\eeq
\beq
\label{6.6bis}
\mu_\text{s}^*=\frac{\frac{d-1}{d}a_2+\zeta^*\kappa^*}{\omega_{2|1}^*}.
\eeq

The expressions of  $\eta_\text{s}^*$, $\kappa_\text{s}^*$ and $\mu_\text{s}^*$ for IHS have been recently derived by considering the first Sonine approximation \cite{MGV14}. Their explicit forms are displayed in the Appendix \ref{appB}. Figure \ref{fig4} shows the dependence of $\eta_\text{s}^*$, $\kappa_\text{s}^*$ and $\mu_\text{s}^*$ on $\al$ for two and three dimensions. We have considered here the theoretical results obtained for Model B of IMM with the power $q=\frac{1}{2}$ (Eqs.\ \eqref{6.4bis}--\eqref{6.6bis}) and the results for IHS (Eqs.\ \eqref{5.24} and \eqref{5.25}). We observe that in general the qualitative dependence of the Navier-Stokes transport coefficients on dissipation of IHS is well captured by IMM. As expected (because the same behavior is observed in analogous systems \cite{S03,MGV14}), the dependence of the transport coefficients on inelasticity is more significant in IMM than in IHS. The quantitative differences between both interaction models increase with inelasticity (especially in the two-dimensional case) and they are much more important in the case of the thermal conductivity than in the cases of the shear viscosity and the coefficient $\mu$. However, and compared with the free cooling case \cite{S03}, the discrepancies found here between IMM and IHS are much less important than those observed in the undriven case.

\section{Conclusions}
\label{sec8}

In this paper, the influence of the interstitial fluid on the dynamic properties of solid particles in a monodisperse suspension has been studied. The fluid-solid interaction force has been modeled via a viscous drag force proportional to the particle velocity. This type of external force has been recently used in different works \cite{H13,SMMD13,WGZS14} to study the shear rheology of frictional hard-sphere suspensions. Our goal here has been to determine the forms of the Navier-Stokes transport coefficients in terms of the relevant parameters of the suspension (coefficients of restitution $\al$ and friction $\gamma$).

To address the above issue in the context of the (inelastic) Boltzmann equation without having to resort to approximate methods or computer simulations, one has to consider simplified collision models. As for elastic collisions \cite{GS03}, the IMM renders itself to an analytical treatment for transport properties since the velocity moments of the Boltzmann collision operator can be \emph{exactly} evaluated without the knowledge of the velocity distribution function. Those collisional moments are given in terms of an effective collision frequency $\nu_0$ independent of the coefficient of restitution $\al$.  As in previous works \cite{SG07}, two different classes of IMM have been studied here: Model A, where $\nu_0$ is independent of temperature, and Model B where $\nu_0$ is an increasing function of temperature ($\nu_0 \propto T^q$).

The Chapman-Enskog method \cite{CC70} has been used to derive Navier-Stokes-order constitutive equations for the momentum and heat fluxes. The results indicate in general a non-negligible influence of the gas phase on the shear viscosity $\eta$, the thermal conductivity $\kappa$ and the coefficient $\mu$ (relating the heat flux with the density gradient). Specifically, the presence of the gas phase lowers $\eta$ and increases $\kappa$ and $\mu$ (see Fig.\ \ref{fig3}). However, for Model B with $q=\frac{1}{2}$, the exact results derived here show that the hydrodynamic forms of $\kappa$ and $\mu$ are \emph{independent} of the friction coefficient $\gamma$. This surprising feature agrees qualitatively well with the previous results derived in Ref.\ \cite{GTSH12} for IHS in the case of the thermal conductivity, since a negligible influence of the gas phase on $\kappa$ was found for this interaction model. With respect to the influence of the initial conditions, our expressions for the heat flux transport coefficients also show that in the case of Model B the presence of the drag force could prevent the existence of hydrodynamic forms for $\kappa$ and $\mu$. The confirmation of this point requires the performance of computer simulations by means of the DSMC method \cite{B94}.

The analysis shows that while the (scaled) zeroth-order distribution function $f^{(0)}$ does not explicitly depend on $\gamma^*$, the transport coefficients associated with the first-order distribution $f^{(1)}$ present in general a complex dependence on the (dimensionless) friction coefficient. This result is fully consistent with previous results \cite{GSB90} derived for ordinary (elastic) gases where it was exactly shown that the effect of the drag force \eqref{5.0} for homogeneous systems of particles interacting via repulsive potentials is just to scale the velocities and to introduce a new time scale. On the other hand, the above scaling fails for inhomogeneous situations (due essentially to the presence of the inhomogeneous term $\mathbf{v}\cdot \nabla f^{(0)}$ in $f^{(1)}$) and the (scaled) transport coefficients are affected by the drag force. The results derived here extend to inelastic systems the conclusions made in Ref.\ \cite{GSB90} since the external force does not play a neutral role for transport and hence, the expressions of the (scaled) transport coefficients obtained with and/or without the drag force are in general different.

Furthermore, the present results generalize to granular flows recent results \cite{PG14} obtained for ordinary gases subjected to a drag force of the form \eqref{5.0}. In this previous work \cite{PG14}, it was assumed for the sake of simplicity that the friction coefficient $\gamma(\mathbf{r},t)\propto \nu(\mathbf{r},t)$ and so,  $\gamma^*\equiv \text{const.}$ The expressions of the Navier-Stokes transport coefficients when $\gamma^*$ is constant can be easily derived by following similar steps as those made here. Their forms are provided in the Appendix \ref{appC} and extend to inelastic collisions ($\al \neq 1$) the results reported in Ref.\ \cite{PG14}.

The knowledge of the Navier-Stokes transport coefficients allows one in principle to solve the linearized hydrodynamic equations around the homogenous time-dependent state (HCS) for solid particles. The determination of the critical length scale $L_c$ in freely cooling flows offers one of the most interesting applications of the Navier-Stokes hydrodynamics and is likely the phenomenon that makes granular flows so different from ordinary gases \cite{GZ93, M93, BLN13, PJDR14}. On the other hand, given that the dimensionless friction coefficient $\gamma^*\propto T(t)^{-q}$ depends on time in our model of suspensions, the determination of $L_c$ is an intricate problem since it requires to numerically solve the corresponding set of differential equations for the hydrodynamic fields. This contrasts with the stability analysis performed recently for driven ordinary gases ($\gamma^*\equiv \text{const.}$) where $L_c$ was analytically determined \cite{PG14}. We plan to perform a linear stability analysis of the Navier-Stokes equations derived in this paper to assess the impact of the surrounding viscous fluid over previous analytical results obtained for dry granular gases \cite{BDKS98,G05}. Another possible direction of study is the extension of the present results for the transport coefficients to the important subject of
polydisperse gas-solid suspensions. Previous works carried out for IMM \cite{G03} have shown the tractability of the Maxwell kinetic theory for these complex systems and stimulate the performance of this study. In particular, given the difficulties associated with multicomponent systems, the tracer limit (a binary mixture where the concentration of one of the species is negligible) could be perhaps a good starting point to provide some insight into the general problem. Work along these lines will be carried out in the near future.

\acknowledgments

We are grateful to Andr\'es Santos for valuable discussions on the subject of this paper. The research of V. G. has been supported by the Spanish Government through grant No. FIS2013-42840-P, partially financed by
FEDER funds and by the Junta de Extremadura (Spain) through Grant No. GR15104.

\appendix
\section{First-order approximation}
\label{appA}

In this Appendix, some technical details of the application of the Chapman-Enskog method to the first-order approximation are provided.  Up to first order in spatial gradients, the velocity distribution function $f^{(1)}(\mathbf{V})$ obeys the kinetic equation
\beq
\label{a1}
\partial_t^{(0)}f^{(1)} -\gamma \frac{\partial}{\partial
{\bf V}}\cdot {\bf V} f^{(1)}+{\cal L}f^{(1)}=-\left(D_t^{(1)}+\mathbf{V}\cdot \nabla \right)f^{(0)}+
\gamma \Delta \mathbf{U}\cdot \frac{\partial f^{(0)}}{\partial
{\bf V}},
\eeq
where the linear operator ${\cal L}f^{(1)}$ is defined by Eq.\ \eqref{5.2} and $D_t^{(1)}\equiv \partial_t^{(1)}+\mathbf{U}\cdot \nabla$. The macroscopic balance equations \eqref{2.7}--\eqref{2.9} to first order in the gradients are
\begin{equation}
\label{a2}
D_t^{(1)}n=-n\nabla\cdot {\bf U},\quad
D_t^{(1)}U_i=-\rho^{-1}\nabla_i p-\gamma \Delta U_i,
\end{equation}
\begin{equation}
\label{a3}
D_t^{(1)}T=-\frac{2T}{d}\nabla\cdot {\bf U}.
\end{equation}
Use of Eqs.\ \eqref{a2} and \eqref{a3} in Eq.\ \eqref{a1} yields
\beq
\label{a4}
\left(\partial_{t}^{(0)}+{\cal L}\right)f^{(1)}-\gamma \frac{\partial}{\partial
{\bf v}}\cdot {\bf V} f^{(1)}=\mathbf{A}\cdot \nabla \ln T+\mathbf{B}\cdot \nabla \ln n +
C_{ij}\frac{1}{2}\left(\nabla_i U_j+\nabla_j U_i-\frac{2}{d}\delta_{ij}\nabla\cdot \mathbf{U}\right),
\eeq
where
\begin{equation}
{\bf A}\left( \mathbf{V}\right)=-\mathbf{V}T\frac{\partial f^{(0)}}{\partial T}-\frac{p}{\rho}\frac{\partial f^{(0)}}{\partial \mathbf{V}},  \label{a5}
\end{equation}
\beq
{\bf B}\left(\mathbf{V}\right)= -{\bf V} f^{(0)}-\frac{p}{\rho}
\frac{\partial f^{(0)}}{\partial \mathbf{V}},  \label{a6}
\eeq
\begin{equation}
\label{a7}
C_{ij}\left(\mathbf{V}\right)=V_i\frac{\partial f^{(0)}}{\partial V_j}.
\end{equation}
Upon deriving Eqs. \eqref{a5}--\eqref{a7}, use has been made of the spherical symmetry of $f^{(0)}(\mathbf{V})$ which allows to express the tensor derivative of the flow field $\nabla_i U_j$ in terms of its independent tracer and traceless parts, namely,
\beqa
\label{a9}
V_i\frac{\partial f^{(0)}}{\partial V_j}\nabla_iU_j&=&\frac{1}{2}V_i\frac{\partial f^{(0)}}{\partial V_j}\left( \nabla_i U_j+\nabla_j U_i\right) \nonumber\\
&=&\frac{1}{2}V_i\frac{\partial f^{(0)}}{\partial V_j}\left( \nabla_i U_j+\nabla_j U_i-\frac{2}{d}\delta_{ij}\nabla \cdot \mathbf{U}\right)+\frac{1}{d}\mathbf{V}\cdot \frac{\partial f^{(0)}}{\partial \mathbf{V}}\nabla \cdot \mathbf{U}.
\eeqa

\section{Inelastic hard spheres results for steady states}
\label{appB}

In this Appendix we give the expressions of the transport coefficients of IHS under steady state conditions ($\gamma^*=-\zeta^*/2$) \cite{GMV13}. For the sake of simplicity and given that the fourth cumulant $a_2$ is very small, we take the Gaussian approximation for the zeroth-order distribution $f^{(0)}$ and hence $a_2=0$. In this case, the expressions of $\eta_\text{s}^*$, $\kappa_\text{s}^*$ and $\mu_\text{s}^*$ are (see Appendix B of Ref.\ \cite{MGV14})
\beq
\label{5.24}
\eta_\text{s}^*=\frac{1}{\nu_\eta^*-\zeta^*},
\eeq
\beq
\label{5.25}
\kappa_\text{s}^*=\frac{d-1}{d}\frac{1}{\nu_\kappa^*-2\zeta^*}, \quad \mu_\text{s}^*=\frac{\zeta^*\kappa^*}{\nu_\kappa^*-\frac{3}{2}\zeta^*},
\eeq
where
\beq
\label{5.27}
\zeta^*=\frac{d+2}{4d}(1-\al^2),
\eeq
\beq
\label{5.28}
\nu_\eta^*=\frac{3}{4d}\left(1-\al+\frac{2}{3}d\right)(1+\al), \quad \nu_\kappa^*=\frac{1+\al}{d}\left[\frac{d-1}{2}+\frac{3}{16}(d+8)(1-\al)\right].
\eeq

\section{Transport coefficients when $\gamma^*$ is constant}
\label{appC}

In this Appendix we provide the explicit expressions of the transport coefficients when the (reduced) friction coefficient $\gamma^*$ is constant. They are given by
\beq
\label{c1}
\eta^*=\frac{1}{\omega_{0|2}^*+q(\zeta^*+2\gamma^*)},
\eeq
\beq
\label{c2}
\kappa^*=\frac{d-1}{d}\frac{1+2a_2}{\omega_{2|1}^*-\left(\frac{1}{2}\zeta^*+\gamma^*\right)}, \quad
\mu^*=\frac{\frac{d-1}{d}a_2+(\zeta^*+2\gamma^*)\kappa^*}{\omega_{2|1}^*+(q-\frac{1}{2})\left(\zeta^*+\gamma^*\right)}.
\eeq
In the case of elastic hard spheres ($\al=1$ and $d=3$), Eqs.\ \eqref{c1}-\eqref{c2} agree with those recently obtained for Maxwell molecules \cite{PG14}.

\end{document}